\documentclass[useAMS]{gGAF2e}
\usepackage{mathrsfs,amsfonts,multirow,float,subfigure}
\pdfoutput=1

\begin{document}


\doi{10.1080/03091920xxxxxxxxx}
 \issn{1029-0419} \issnp{0309-1929} \jvol{00} \jnum{00} \jyear{2009} 

\markboth{Robert J. Teed, Chris A. Jones \& Rainer Hollerbach}{Rotating convection with zonal flow}

\title{Rapidly rotating plane layer convection with zonal flow}

\author{Robert J. Teed, Chris A. Jones \& Rainer Hollerbach, University of Leeds, UK}

\maketitle

\begin{abstract}
The onset of convection in a rapidly rotating layer in which a thermal wind 
is present is studied. Diffusive effects are included. The main motivation
is from convection in planetary interiors, where thermal winds are expected
due to temperature variations on the core-mantle boundary. The system admits both
convective instability and baroclinic instability. We find a smooth transition
between the two types of modes, and investigate where the
transition region between the two types of instability occurs in parameter
space. The thermal wind helps to destabilise the convective modes.
Baroclinic instability can occur when the 
applied vertical temperature gradient is stable, and the critical Rayleigh
number is then negative. Long wavelength modes are the first to become unstable.
Asymptotic analysis is possible for the transition region and also for long
wavelength instabilities, and the results agree well with our numerical solutions.
We also investigate how the instabilities in this system relate to the classical
baroclinic instability in the Eady problem. We conclude by noting that baroclinic
instabilities in the Earth's core arising from heterogeneity in the lower mantle
 could possibly drive a dynamo even if the Earth's core were stably stratified
and so not convecting.

\bigskip

\bigskip

\end{abstract}


\section{Introduction}

The geomagnetic field is believed to be generated by convection in the Earth's fluid outer core. The 
convection in the core
is strongly influenced by rotation, leading to the formation of tall thin columns which transport
the heat out from the interior to the core-mantle boundary \citep{buscar76,jon00}. 
The form of these columns plays
a vital role in the mechanism by which magnetic field is generated \citep{Ols99}. Although
the convection in the core is in a strongly nonlinear regime, with Rayleigh number well above
that at the onset of convection, dynamo models show that the convecting columns still have many 
features that resemble the pattern of convection derived from linear theory.    

The onset of convection from stationary fluid in rapidly rotating spherical bodies is 
now fairly well understood.
\citet{rob68} and \citet{bus70} evaluated the essential principles, confirmed in numerical
studies by \citet{zha92}. The behaviour in the asymptotic limit of small Ekman number
(rapid rotation) was elucidated by \citet{jonsow00} and \citet{dor04}. 
In this paper we study the onset of convection in a rotating system with an imposed zonal flow,
that is an axisymmetric, azimuthal flow.
Zonal flows occur frequently in nature. Well-known examples are the wind systems on the giant
planets, where east-west flows reaching up to several hundreds of metres per second can occur.
The systems most relevant to this paper are the cases where the zonal flow is a thermal wind,
driven by latitudinal temperature gradients. A famous example is the jet-stream in our atmosphere,
driven by the pole-equator temperature difference. Thermal winds are also believed to occur
in the Earth's core \citep{olsaur99,sre05,sre06} where warmer
regions above the poles lead to anticyclonic vortices which can be detected in the secular
variation as the geomagnetic field is advected by the flow. This process has been modelled
in the laboratory by \citet{aur03}.

Convection in the outer core is significantly affected by the presence of a solid inner
core of radius approximately 0.35 times the radius of the fluid outer core. 
The fundamental cause of the warmer (and compositionally
lighter) regions near the poles is believed to be the different efficiency of convection
in the polar regions inside the tangent cylinder and outside the tangent cylinder
\citep{til97}. The tangent
cylinder is the imaginary cylinder that touches the inner core; outside the tangent cylinder 
convection columns can reach right across the outer core, but inside the tangent cylinder
columns are bounded by the inner core. Thermal winds inside the Earth's core could also arise
more directly, because of a heterogeneous heat flux across the core-mantle boundary.
Seismic tomography suggests that heterogeneities exist, and a natural interpretation
is that the variations in seismic velocity are due to thermal variations
caused by a core-mantle heat flux that varies with latitude and longitude \citep{gub07}.
In this situation, even when the temperature gradient is subadiabatic so convection would
not be expected, a basic state with no flow is impossible \citep[see e.g.][]{zha96}.
A thermal wind is set up which might lead to a baroclinic instability. The possibility
that the core is stably stratified just below the core-mantle boundary was originally suggested
by \citet{bra93}. While we do not currently know whether the core heat flux is low enough
for such a subadiabatic region to exist, estimates of the thermal conditions in the
core suggest it is a realistic possibility \citep{anu05}. 

The aim of this paper is to examine the effect of a thermal wind on the onset of convection,
and to examine whether baroclinic instabilities can arise in rapidly rotating systems
when the fluid is stably stratified. As we see below, as the thermal wind is gradually increased,
convective modes evolve into baroclinic modes. The critical Rayleigh number can therefore become negative
when the thermal wind flow is large enough that baroclinicity becomes important. This
can occur at conditions which are realistic for the core. To elucidate the fundamental
mechanisms involved, we consider here a simple plane layer model, which allows some asymptotic
limits to be explored. This simple model is most relevant to the polar regions
in the core, since we are taking gravity and rotation to be parallel. More realistic geometries
for core convection will be explored subsequently.    

The onset of rotating convection in a plane layer in the absence of a thermal wind was comprehensively
studied by \citet{cha61}. Baroclinic instability in a stably stratified layer forms the basis
of the Eady problem, discussed in detail in the meteorological context by \citet{ped87} and \citet{drarei81}. 
Here we combine these two classical problems by examining the stability of a simple thermal wind state
when diffusion is present and when the vertical temperature gradient, specified by the Rayleigh number,
can be either positive or negative.


\section{Description of the model}

We consider a plane layer of depth $d$ rotating about the vertical axis with angular velocity
$\Omega$. We choose a Cartesian coordinate system with the origin 
situated at the centre of the layer so that the boundaries are located at $z=\pm d/2$. 
In this geometry $x$ and $y$ are playing the role of the azimuthal and latitudinal coordinates respectively. 
The static temperature gradient in the absence of the zonal flow is such that 
$T=\beta d$  and $T=0$ at $z=-d/2$ and $z=d/2$ respectively. Gravity, $g$, acts downwards 
in the negative $z$-direction. This type of setup is appropriate 
for polar regions of the Earth's core where gravity is near parallel to the rotation axis and the 
zonal flows are expected to depend on $z$.

The equation of motion in a rotating frame whilst assuming the Boussinesq approximation is
\begin{eqnarray}
\frac{\partial\mathbf{U}}{\partial t} + (\mathbf{U}\cdot\nabla)\mathbf{U}+2\Omega\mathbf{\hat{z}}
\times\mathbf{U} &=& -\frac{1}{\rho_0}\nabla \mathcal{P} + g \alpha T\mathbf{\hat{z}} + \nu\nabla^2\mathbf{U}, 
\label{eq:ns}
\end{eqnarray}
and the temperature equation is
\begin{eqnarray}
\frac{\partial T}{\partial t} + (\mathbf{U}\cdot\nabla)T &=& \kappa\nabla^2T, \label{eq:temp}
\end{eqnarray}
where $\alpha$, $\nu$ and $\kappa$ are the coefficient of thermal expansion, the kinematic 
viscosity and the thermal diffusivity respectively, $\mathcal{P}$ being the pressure and $\rho_0$ the
density. Also, the Boussinesq continuity equation is simply
\begin{eqnarray}
\nabla\cdot\mathbf{U} = 0 \label{eq:cont}.
\end{eqnarray}

\subsection{Basic state}\label{sec:basic}

In many models the basic state has a velocity field set to zero and we have hydrostatic balance in 
the momentum equation between the pressure gradient and the buoyancy. When this is the case taking 
the curl of (\ref{eq:ns}) results in a $T$ that can only vary in the direction parallel to gravity. 
However if the basic state temperature varies in the $x$ or $y$ direction we must have a balance 
between the pressure gradient, buoyancy and Coriolis force in the momentum equation. By taking the 
curl of (\ref{eq:ns}) in this case we obtain the thermal wind equation
\begin{equation}
2\Omega\frac{\partial\mathbf{U}}{\partial z} = g \alpha \mathbf{\hat{z}}\times\nabla T, \label{eq:tw}
\end{equation}
which generates an azimuthal zonal flow, the thermal wind, when $T$ has $y$-dependence.

Since we want a thermal wind in our basic state, we set $\mathbf{U}$, $T$ and $\mathcal{P}$ as 
$\mathbf{u_0}$, $T_0$ and $p_0$ respectively, and let
\begin{eqnarray}
\mathbf{u_0} &=& u^\prime z\mathbf{\hat{x}}, \label{eq:ubasic} \\
T_0 &=& \beta \left( \frac{d}{2}-z \right) - \frac{2\Omega u^\prime }{ g \alpha} y, \\
p_0 &=& g \alpha  \beta \rho_0 \left( \frac{zd}{2}-\frac{z^2}{2} \right) 
- 2\rho_0 \Omega u^\prime y z + p_{\textrm{constant}},
\end{eqnarray}
which is a solution to the system of equations (\ref{eq:ns}) - (\ref{eq:tw}), where $\beta$ is the static temperature gradient in the absence of the zonal flow. Here $u^\prime$ is the constant shear defining the strength of the zonal flow.
These equations define the basic state. Of particular note here is the fact that the temperature distribution depends on a coordinate other than the coordinate parallel to the rotation axis, so the basic state is baroclinic, that is $\nabla p_0$ is
not parallel to $\nabla \rho = - \alpha \rho_0 \nabla T_0$.

\subsection{Perturbed state}

In order to analyse linear stability we now add small perturbations to the basic state so that $\mathbf{U}=\mathbf{u_0}+\mathbf{u}$, $\mathcal{P} = p_0 + p$ and $T=T_0+\theta$. Since the perturbations are small we are able to ignore nonlinear terms so that equations (\ref{eq:ns}) and (\ref{eq:temp}), using the definition of the basic state, give
\begin{gather}
\frac{\partial\mathbf{u}}{\partial t} + u^\prime z\frac{\partial\mathbf{u}}{\partial x} + u^\prime u_z \mathbf{\hat{x}} + 2\Omega\mathbf{\hat{z}}\times\mathbf{u} = -\frac{1}{\rho_0}\nabla p + g \alpha \theta\mathbf{\hat{z}} + 
\nu\nabla^2\mathbf{u}, \label{eq:nsptb2} \\ 
\frac{\partial\theta}{\partial t} + u^\prime z\frac{\partial\theta}{\partial x} - \beta u_z - 
\frac{2\Omega u^\prime}{g \alpha } u_y = \kappa\nabla^2\theta. \label{eq:tempptb2}
\end{gather}

We proceed by eliminating the pressure to leave four equations for four unknowns. We denote the vorticity by 
{\boldmath$\omega$}  and then the $z$-components of the curl and double curl of equation (\ref{eq:nsptb2}) are
\begin{gather}
\frac{\partial\omega_z}{\partial t} + u^\prime z\frac{\partial\omega_z}{\partial x} - u^\prime\frac{\partial u_z}{\partial y} - 2\Omega\frac{\partial u_z}{\partial z} = \nu\nabla^2\omega_z, \label{eq:vort1} \\
\frac{\partial\nabla^2 u_z}{\partial t} + u^\prime z\frac{\partial \nabla^2 u_z}{\partial x} + 2\Omega\frac{\partial\omega_z}{\partial z} = g \alpha \nabla_H^2\theta + \nu\nabla^4 u_z, \label{eq:curlvort}
\end{gather}
respectively. Here $\nabla_H^2 = \partial^2/\partial x^2 + \partial^2/\partial y^2$ is the horizontal Laplacian. Then by employing the identity $\partial\omega_z/\partial x - \partial^2u_z/\partial y\partial z = \nabla_H^2 u_y$, equation (\ref{eq:tempptb2}) can be written
\begin{eqnarray}
\nabla_H^2\left(\frac{\partial\theta}{\partial t} + u^\prime z\frac{\partial\theta}{\partial x} - \beta u_z - \kappa\nabla^2\theta\right) = \frac{2\Omega u^\prime}{g \alpha }\left(\frac{\partial\omega_z}{\partial x}-\frac{\partial^2u_z}{\partial y\partial z}\right). \label{eq:heat2}
\end{eqnarray}
We now have three equations (\ref{eq:vort1}) - (\ref{eq:heat2}) for three unknowns, namely: $u_z$, $\omega_z$ and $\theta$. Next we non-dimensionalise these equations using length scale $d$, time scale $d^2/\nu$ and temperature scale $\beta\nu d/\kappa$. Then equations (\ref{eq:vort1}) - (\ref{eq:heat2}) become
\begin{gather}
\left(\frac{\partial}{\partial t} + Re z\frac{\partial}{\partial x} - \nabla^2\right)\omega_z - Re\frac{\partial u_z}{\partial y} - E^{-1}\frac{\partial u_z}{\partial z} = 0, \label{eq:finEvort} \\
\left(\frac{\partial}{\partial t} + Re z\frac{\partial}{\partial x} - \nabla^2\right)\nabla^2u_z + E^{-1}\frac{\partial\omega_z}{\partial z} = Ra\nabla_H^2\theta, \\
P\left(\frac{\partial}{\partial t} + Re z\frac{\partial}{\partial x} - P^{-1}\nabla^2\right)\nabla_H^2\theta = \nabla_H^2u_z + \frac{PRe}{ERa}\left(\frac{\partial\omega_z}{\partial x} - \frac{\partial^2u_z}{\partial y\partial z}\right), \label{eq:finEtemp}
\end{gather}
where the Ekman number, $E$, Prandtl number, $P$, Rayleigh number, $Ra$, and Reynolds number, $Re$, are defined as
\begin{equation}
E=\frac{\nu}{2\Omega d^2}, \qquad P=\frac{\nu}{\kappa}, \qquad Ra=\frac{g\alpha\beta d^4}{\nu\kappa}, \qquad Re=\frac{u' d^2}{\nu}.
\end{equation}
Equations (\ref{eq:finEvort}) - (\ref{eq:finEtemp}) are the finite Ekman number equations for rapidly rotating plane layer convection with zonal flow. Our system is defined so that when $\beta>0$ we have cold fluid sitting on top of hot fluid and thus the layer is buoyantly unstable. Therefore, as is usually the case when considering thermal convection, we require a positive Rayleigh number above some critical value, $Ra_c$, for convective motions to begin. In the case where $\beta<0$ the system is buoyantly stable since hot fluid sits on top of cold fluid and with a basic state temperature distribution only dependent on $z$ no convection is possible. However, since the basic state temperature distribution we have defined in section \ref{sec:basic} depends on $y$ as well as $z$ it is not immediately clear if motion is forbidden when $Ra<0$ in our setup.


\section{Numerics}

The solutions were assumed to take the form: $\exp(\sigma t+\mathrm{i}(k_xx+k_yy))$ where the growth rate, $\sigma$, is in general, complex. The resulting equations are
\begin{gather}
\left(\sigma + \mathrm{i}k_x Re z + k^2 -\frac{\mathrm{d}^2}{\mathrm{d}z^2}\right)\omega_z - \mathrm{i}k_y Re u_z - E^{-1}\frac{\mathrm{d}u_z}{\mathrm{d}z} = 0,
\label{eq:vortbc} \\
\left(\sigma + \mathrm{i}k_x Re z + k^2 - \frac{\mathrm{d}^2}{\mathrm{d}z^2}\right)\left(\frac{\mathrm{d}^2}{\mathrm{d}z^2}-k^2\right)u_z + E^{-1}\frac{\mathrm{d}\omega_z}{\mathrm{d}z} = -k^2 Ra \theta,
\label{eq:curlvortbc} \\
\left(\sigma P + \mathrm{i}k_x P Re z + k^2 - \frac{\mathrm{d}^2}{\mathrm{d}z^2}\right)\theta = u_z - 
\frac{\mathrm{i}P Re}{E Ra k^2}\left(k_x\omega_z-k_y\frac{\mathrm{d}u_z}{\mathrm{d}z}\right),
\label{eq:tempbc}
\end{gather}
where $k^2=k_x^2+k_y^2$. In addition to demanding that there be no penetration ($u_z$=0) and a 
constant surface temperature ($\theta$=0) at the boundaries, we considered two cases, namely 
stress-free and no-slip boundary conditions on both the upper and lower boundaries so that
\begin{align}
\frac{\mathrm{d}^2u_z}{\mathrm{d}z^2} =& 0 = \frac{\mathrm{d}\omega_z}{\mathrm{d}z}\ \textrm{at}\ z=\pm\frac{1}{2}\ \textrm{for the 
stress-free case,}  \label{eq:sfbcs} \\
\frac{\mathrm{d}u_z}{\mathrm{d}z} =& 0 = \omega_z\ \textrm{at}\ z=\pm\frac{1}{2}\ \textrm{for the no-slip case.} \label{eq:nsbcs}
\end{align}
We solved equations (\ref{eq:vortbc}) - (\ref{eq:tempbc}) using a simple eigenvalue solver. The system has 
the following six input parameters: $k_x$, $k_y$, $Re$, $P$, $E$ and $Ra$, which can be varied to obtain 
the growth rate. Given values for the other five parameters we are interested in finding the Rayleigh number, $Ra^*$,
at the onset of convection. Hence for various values of the input parameters we searched for marginal modes, 
where $\Re[\sigma]=0$, and recorded the value of $Ra^*$ for which the mode appeared. To 
reduce the parameter space we worked with typical values of the Ekman number $(E\sim 10^{-3}-10^{-5})$ and 
Prandtl number $(P\sim 0.1-10)$. 

\begin{figure}[t]
\centering
\hspace{-20pt}\subfigure[Stress-free boundaries.]{\label{fig:Raressf}\includegraphics[width=0.50\textwidth,angle=0]{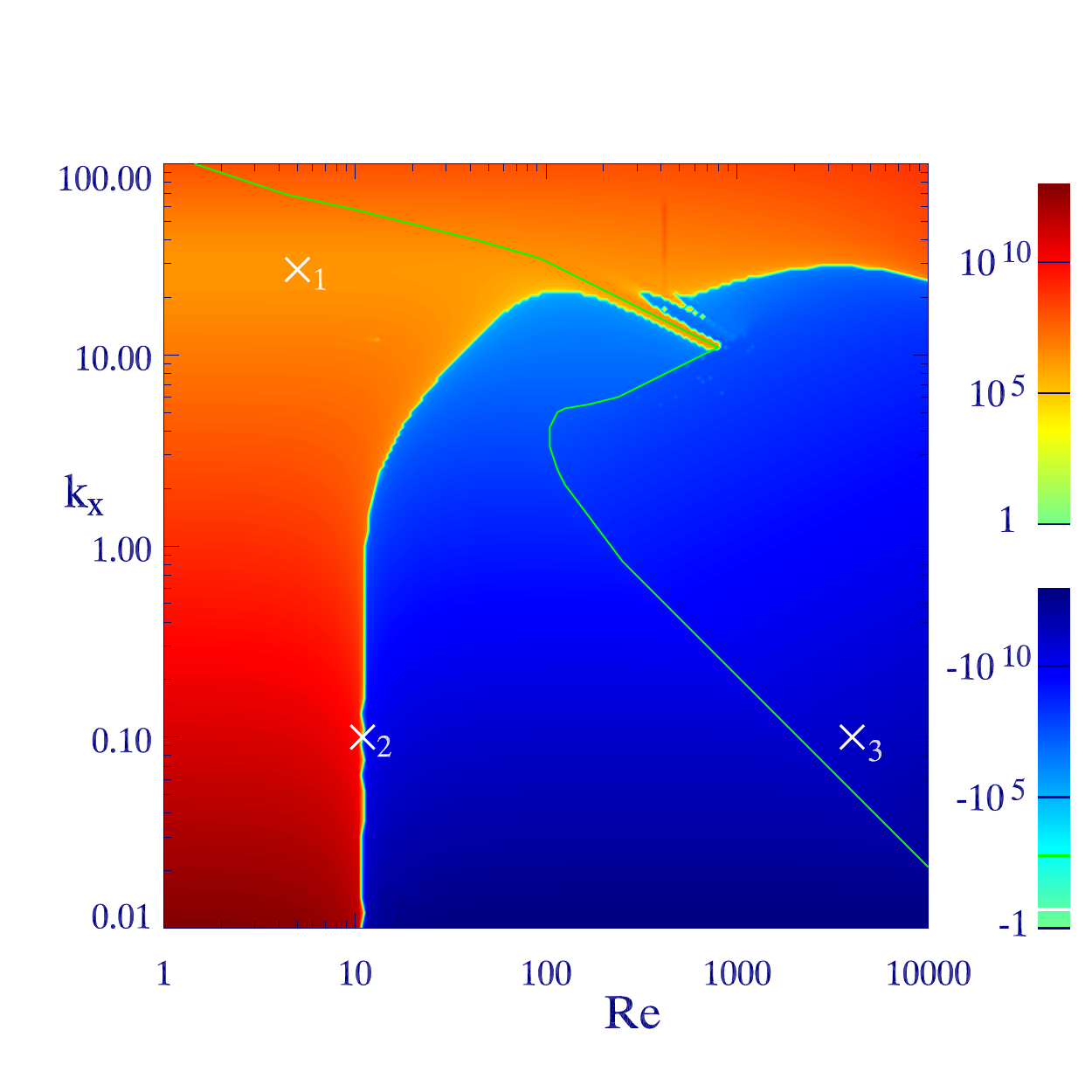}}
\subfigure[No-slip boundaries.]{\label{fig:Raresns}\includegraphics[width=0.50\textwidth,angle=0]{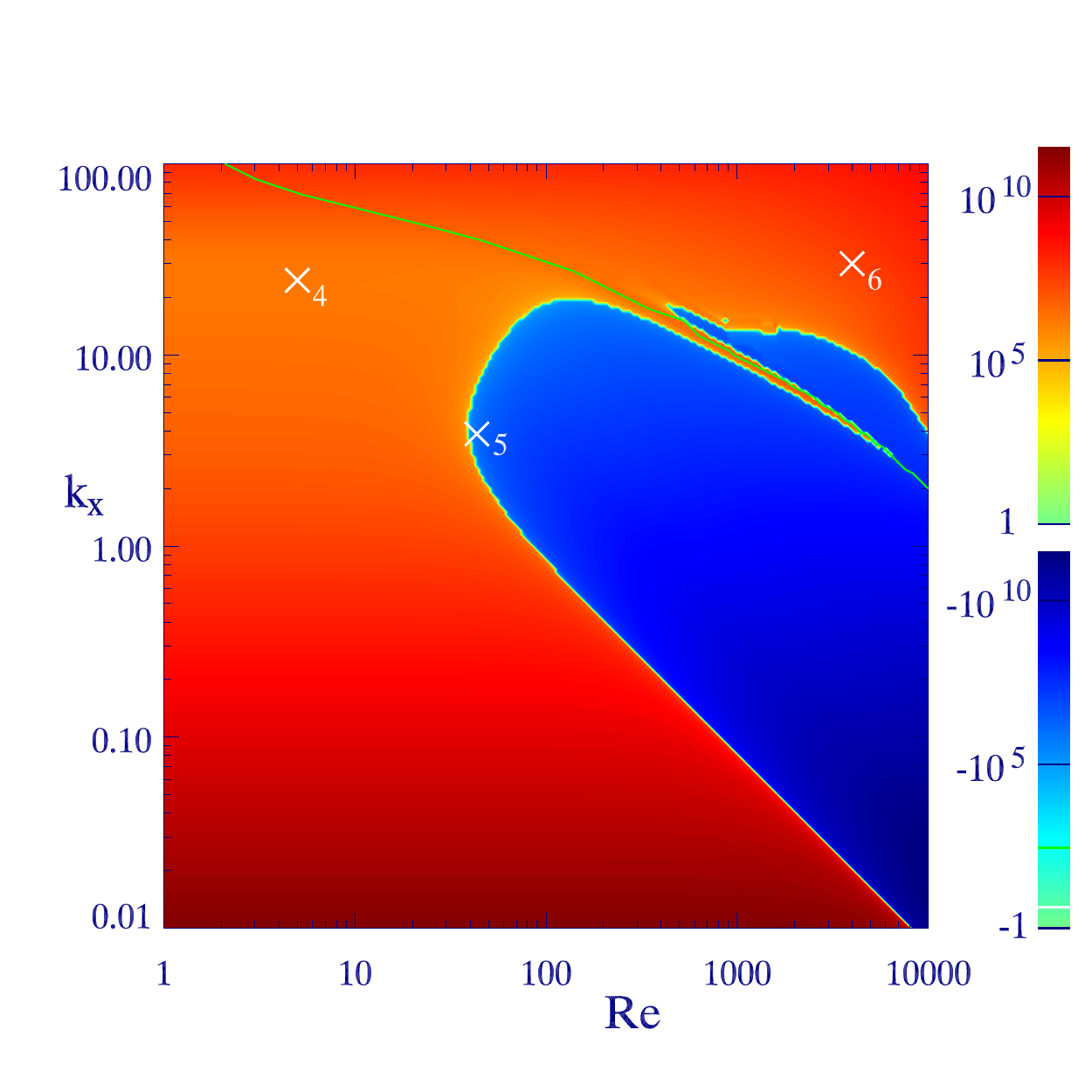}}
\caption{Contour plots of the numerical results for the Rayleigh 
number at onset for $Re$ against $k_x$ with 
$E=10^{-4}$, $P=1$, $k_y=k_{y_c}=0$. The colour scales denote the value of the Rayleigh
number at onset, $Ra^*$.
The green curves divide the regions of steady modes and oscillatory modes, onset being
oscillatory to the right of these curves. } \label{fig:Rares}
\end{figure}

\begin{figure}[p]
\centering
\hspace{-50pt}
\subfigure[$\times_1$]{\label{fig:x1}\includegraphics[width=0.4\textwidth]{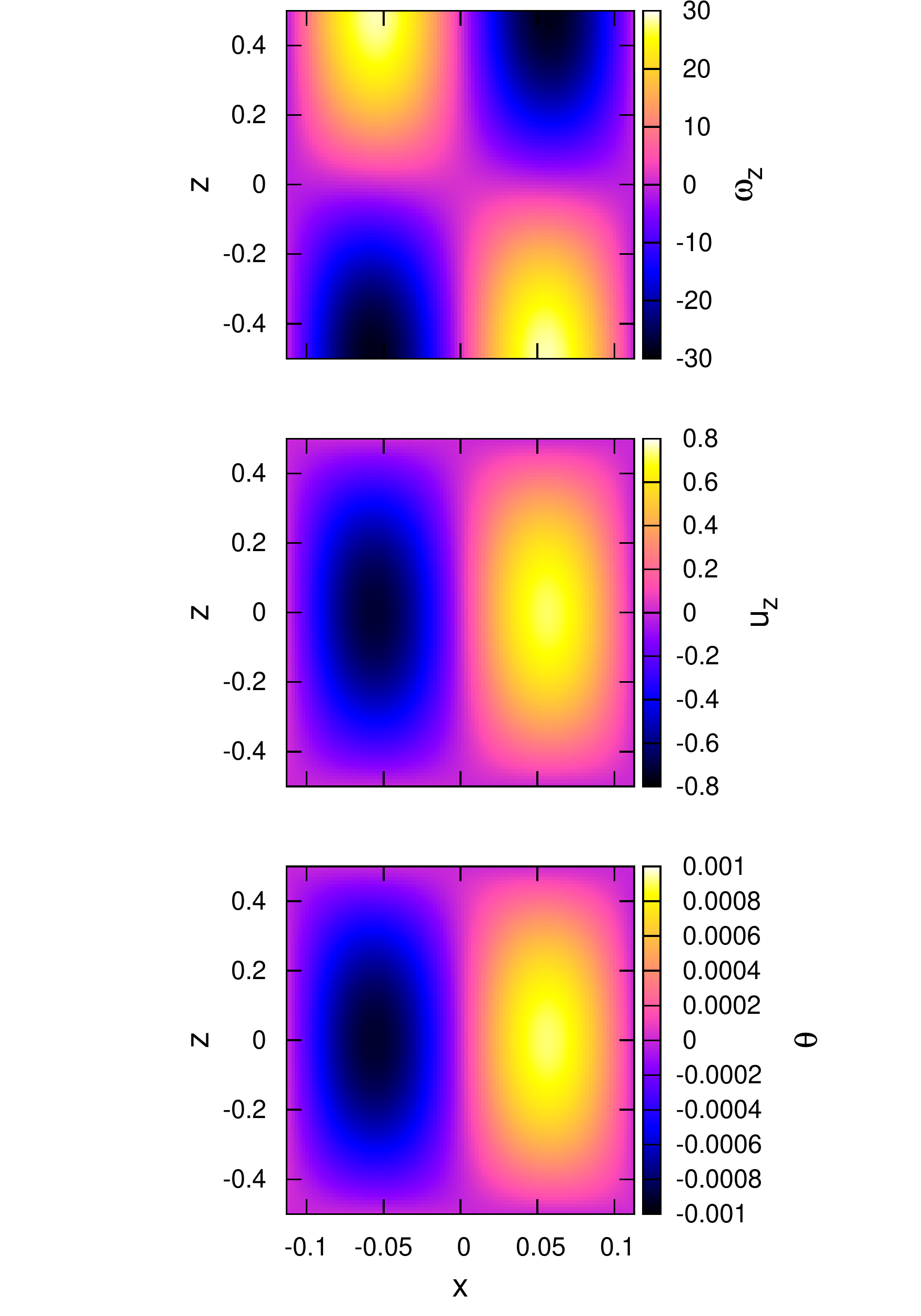}}
\hspace{-50pt}
\subfigure[$\times_2$]{\label{fig:x2}\includegraphics[width=0.4\textwidth]{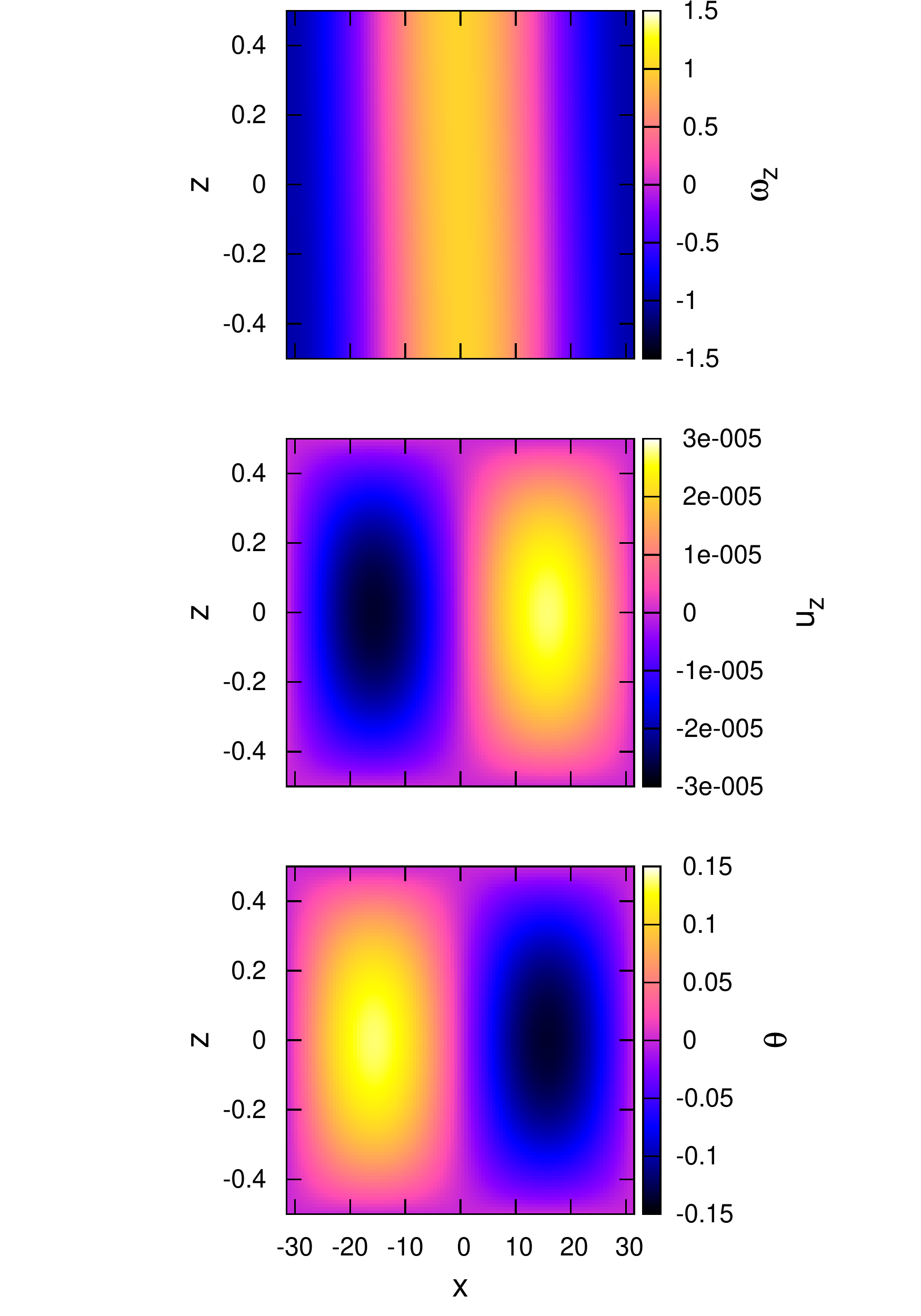}}
\hspace{-50pt}
\subfigure[$\times_3$]{\label{fig:x3}\includegraphics[width=0.4\textwidth]{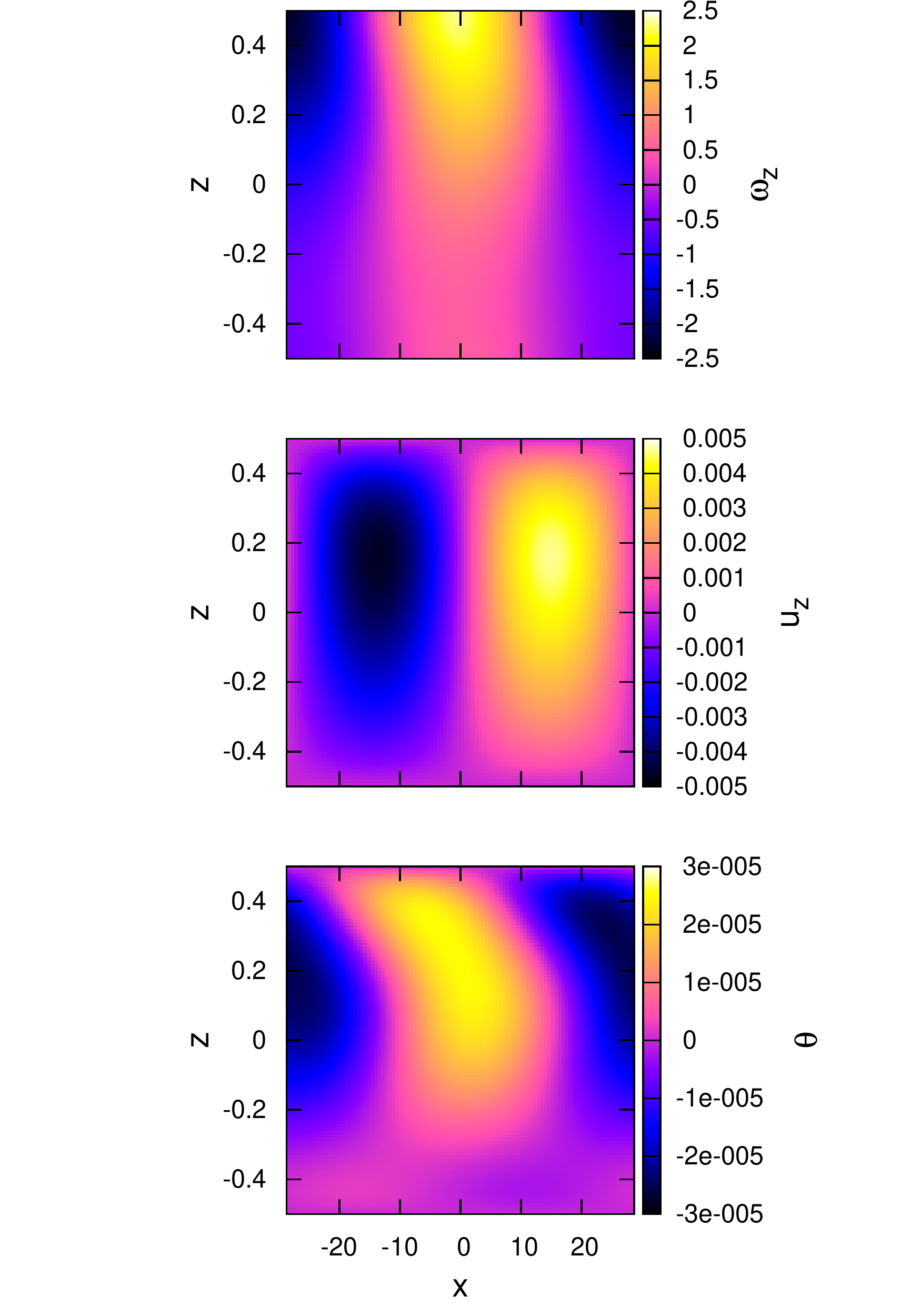}}

\vspace{18pt}

\hspace{-50pt}
\subfigure[$\times_4$]{\label{fig:x4}\includegraphics[width=0.4\textwidth]{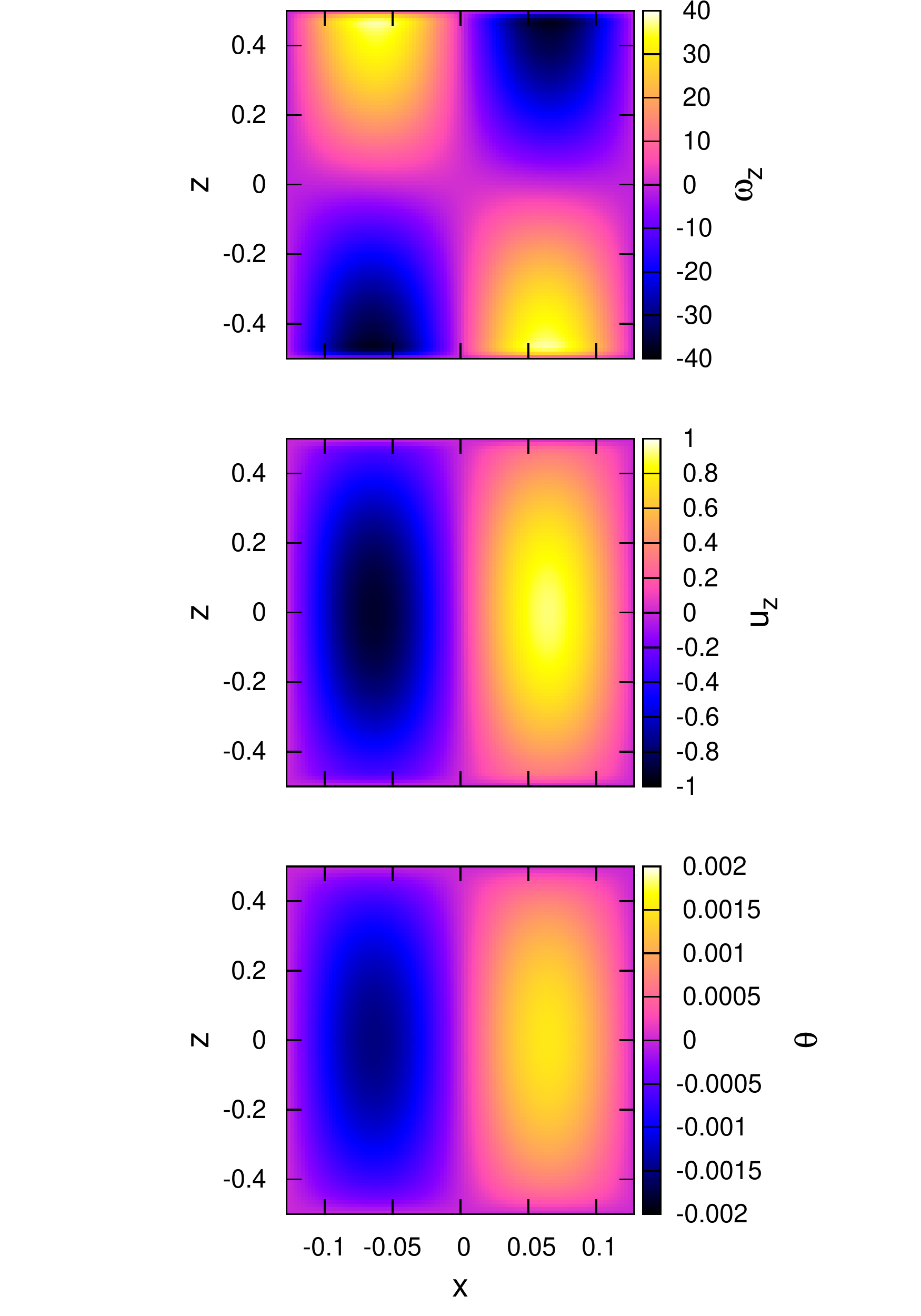}}
\hspace{-50pt}
\subfigure[$\times_5$]{\label{fig:x5}\includegraphics[width=0.4\textwidth]{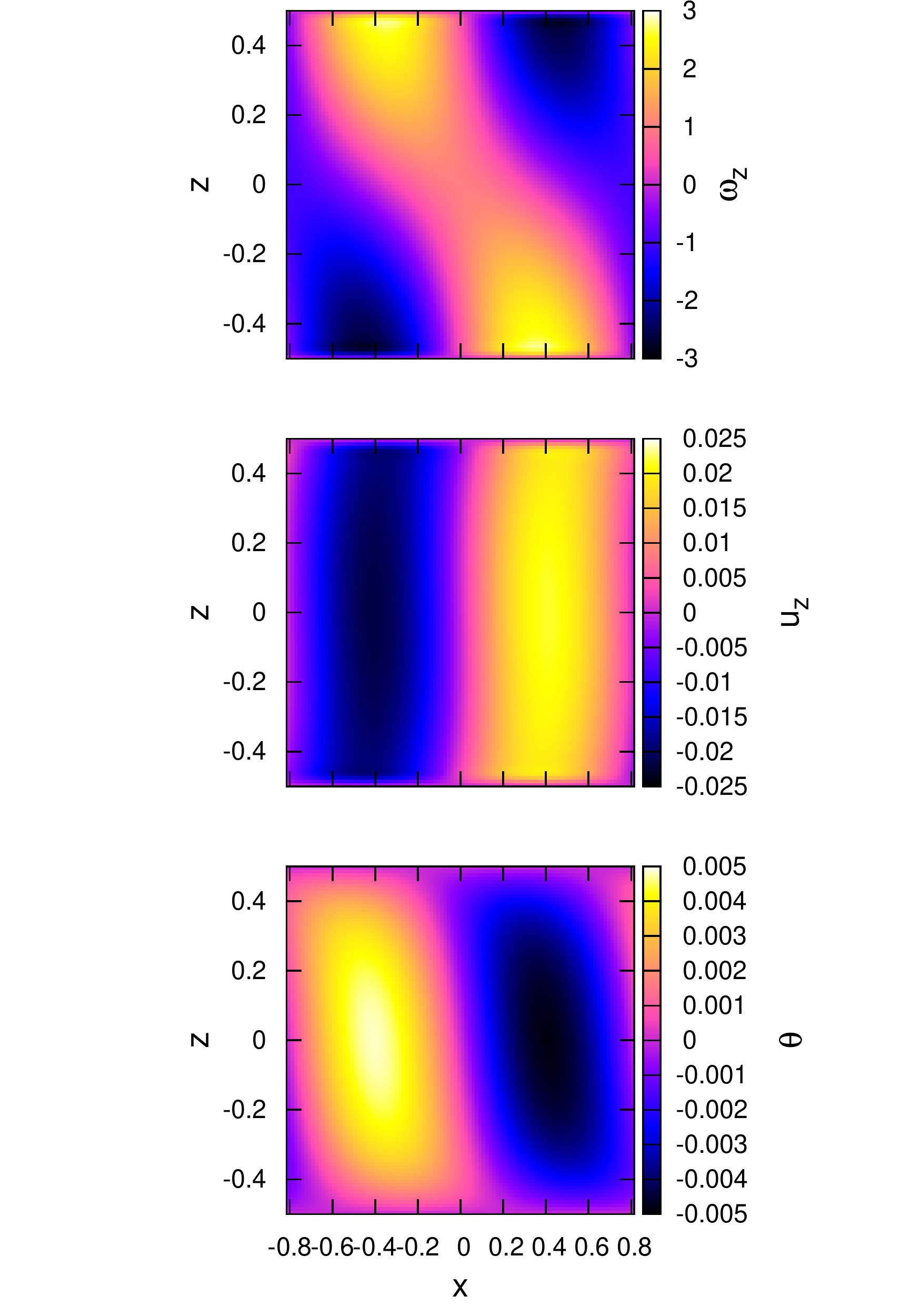}}
\hspace{-50pt}
\subfigure[$\times_6$]{\label{fig:x6}\includegraphics[width=0.4\textwidth]{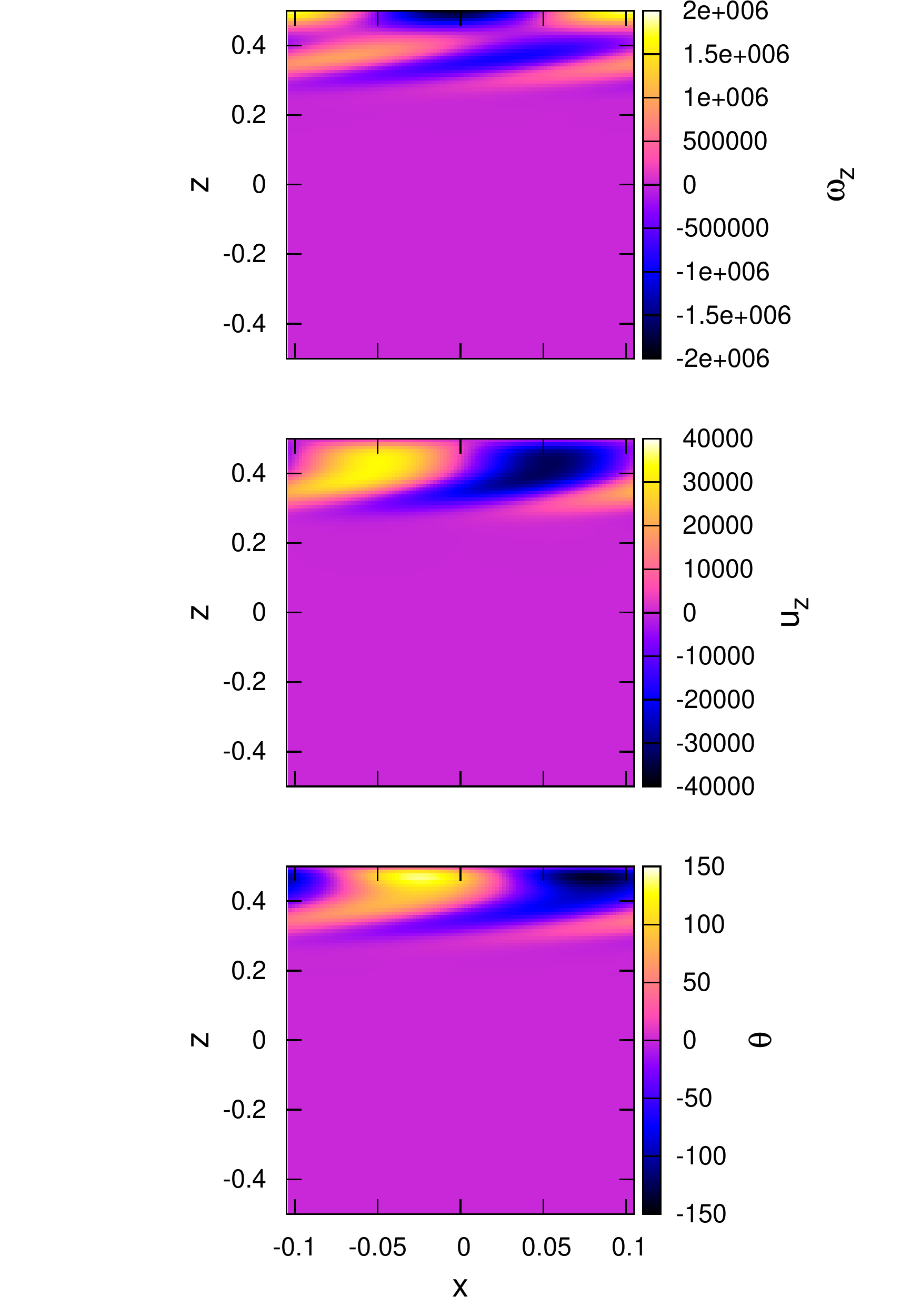}}

\vspace{15pt}

\caption{Eigenfunction plots corresponding to points marked on figure \ref{fig:Rares}. 
 Stress-free cases: $\times_1: Re=5,\ Ra=Ra_c\equiv1.8889\times10^6,\ k_x=k_{x_c}\equiv27.9610.\ 
\times_2: Re=Re^*\equiv10.9599,\ Ra=-10^6,\ k_x=0.1.\ 
\times_3: Re=4000,\ Ra=Ra^*\equiv-1.3571\times 10^{11},\ k_x=0.1.\ $
No-slip cases: $\times_4: Re=5,\ Ra=Ra_c\equiv1.5193\times 10^6,\ k_x=k_{x_c}\equiv24.5630.\ 
\times_5: Re=Re_c\equiv43.4458,\ Ra=-10^6,\ k_x=k_{x_c}\equiv3.8551.\ 
\times_6: Re=4000,\ Ra=Ra^*\equiv3.1259\times 10^7,\ k_x=30.$ $k_y=k_{y_c}=0$ for all points.}
\label{fig:efcnssf}
\end{figure}

Figure \ref{fig:Rares} shows how the onset of convection changes as the azimuthal wavenumber and the 
zonal wind are varied for a particular choice of the Ekman number, Prandtl number and the latitudinal 
wavenumber, for both choices of boundary conditions. It should be noted that the data in figure 
\ref{fig:Rares} is represented on a log-log plot due to the varying magnitudes involved, and a log scale
is necessary for the values of $Ra^*$ also. Since we have positive and negative Rayleigh numbers,
we plot only contours with $\vert Ra^* \vert >1$, but this excludes only a tiny region in figures 1(a) and
1(b). Also of note is the fact that the quantity which has been plotted, $Ra^*$, is not the same as the 
critical Rayleigh number, $Ra_c$, since the latter is minimised over the wavenumbers, $k_x$ and $k_y$. We plot $Ra^*$ here
rather than the critical Rayleigh number due to reasons discussed in section \ref{sec:barreg}. Plots for $Ra_c$ are displayed later. 
The initial striking feature of both sets of results is the appearance of marginal modes with negative Rayleigh number. 
We see that these modes only appear under certain parameter regimes, namely for sufficiently large $Re$ 
and sufficiently small $k_x$. Hence we are able to divide the parameter space into two regimes driven by different types of instability: the convective regime and the baroclinic regime. In the convective/baroclinic regime it is the buoyancy/shear, which is driving the instability. The form of the eigenfunctions in $xz$-space for the points marked in 
figure \ref{fig:Rares} is shown in figure \ref{fig:efcnssf}.

\begin{figure}[t]
\centering
\hspace{-10pt}\subfigure[Stress-free boundaries.]{\label{fig:Racrit}\includegraphics[width=0.50\textwidth]{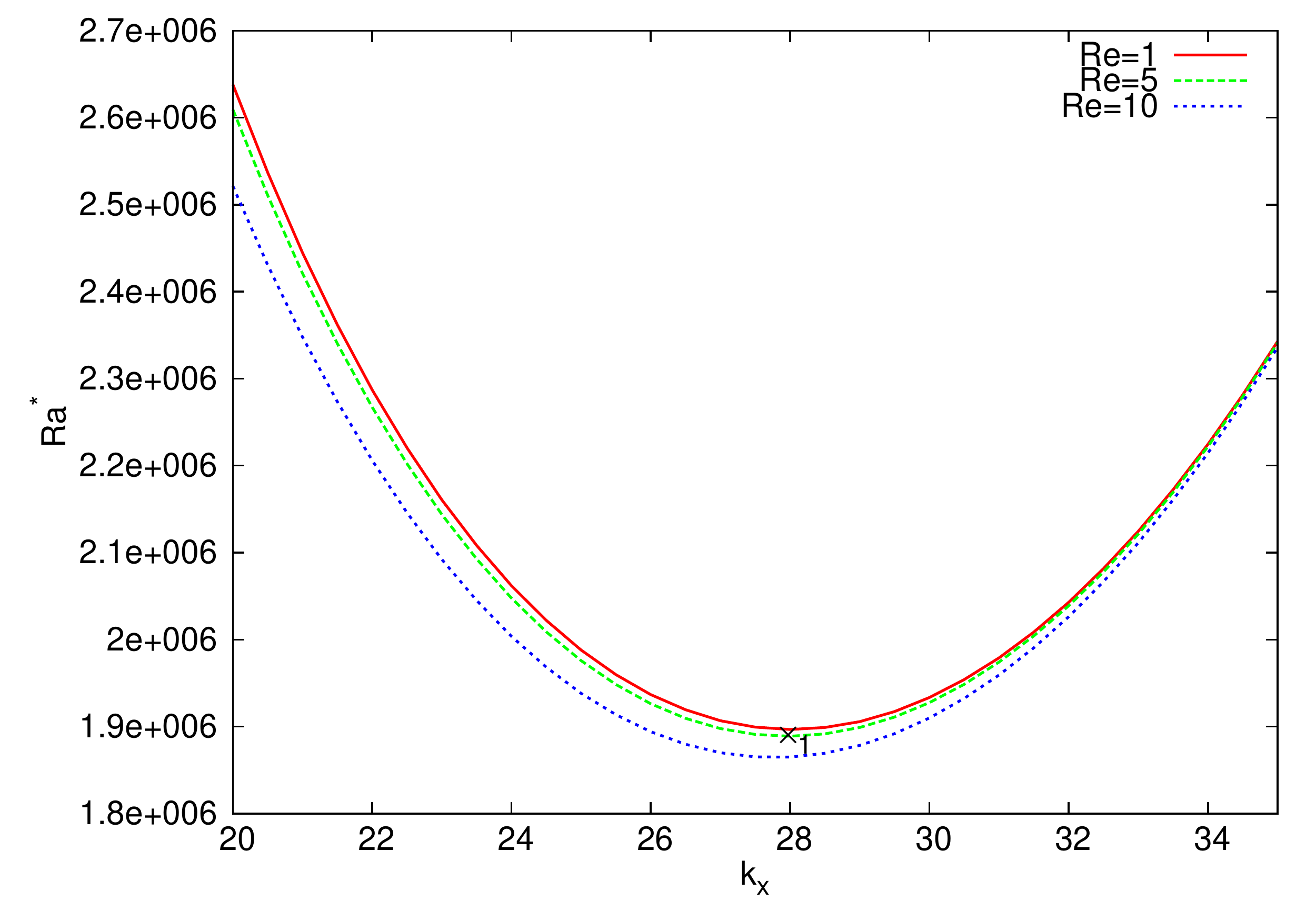}}
\subfigure[No-slip boundaries.]{\label{fig:Racrit2}\includegraphics[width=0.50\textwidth]{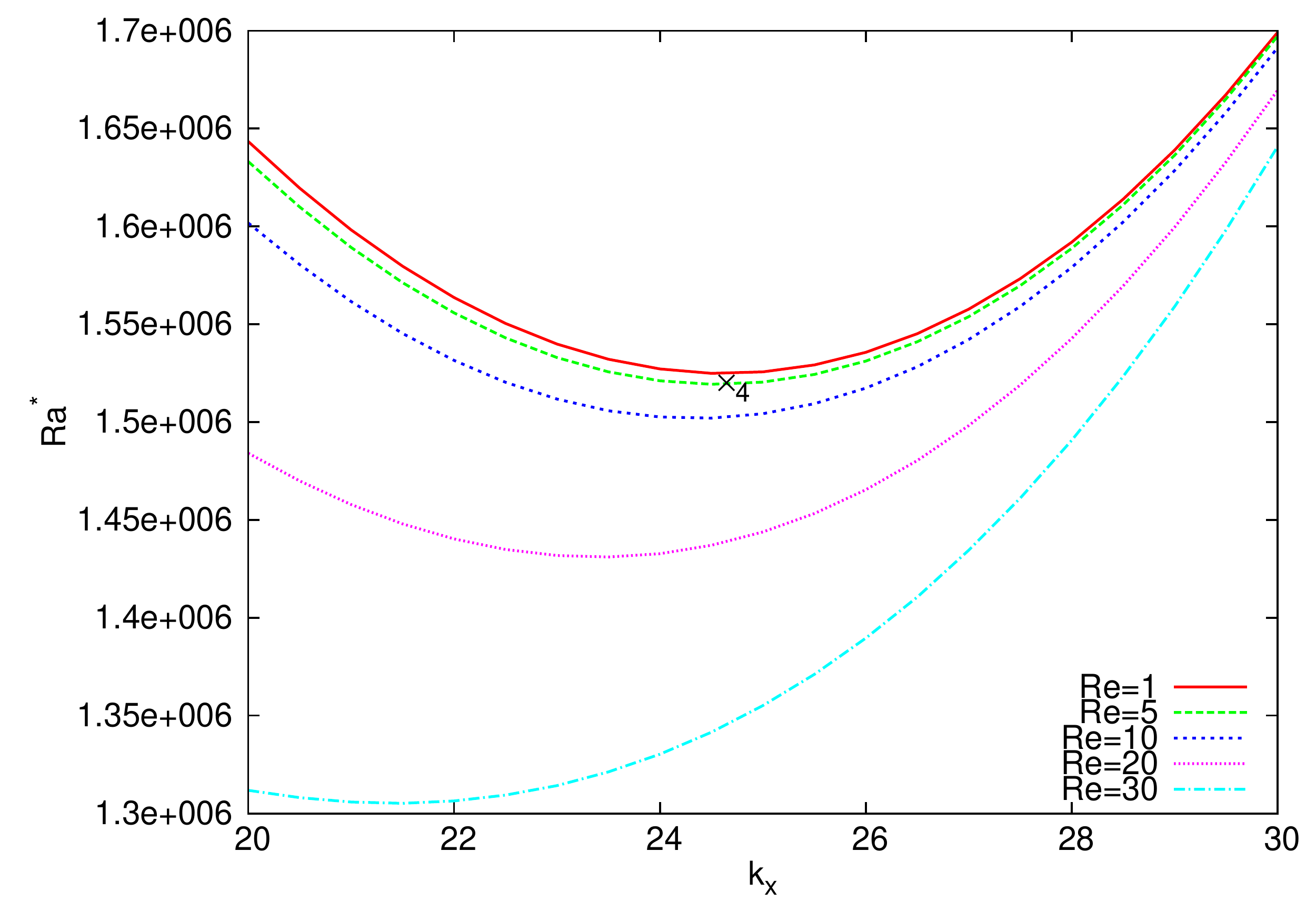}}
\caption{Plots of the numerical results for the onset parameters in the convective regime against $k_x$ with 
$E=10^{-4}$, $P=1$, $k_y=k_{y_c}=0$. The onset parameter is the Rayleigh number in the convective regime.}
\label{fig:Racrits}
\end{figure}

\subsection{Convective regime} \label{sec:conreg}

For low values of the zonal wind we expect to find the usual convective columnar roll solutions 
as described by \citet{cha61}, which we refer to as the `convective modes'. Convective 
modes with the $z$-vorticity antisymmetric about the equator are expected as the most unstable 
modes in plane layer convection; the converse is true in the case of the full sphere as originally 
noted by \citet{bus70}. Indeed for the point marked $\times_1$ we find the mode to be of this form, 
as shown by figure \ref{fig:x1}. The structure has tall thin cells with hot fluid rising and cold 
fluid sinking as expected. This is the case for both types of boundary conditions as is evident from the similarity of figure \ref{fig:x4}, point $\times_4$, for the no-slip case. We also note that for $Re=0$ if we minimise the Rayleigh number at onset over $k$, 
to find the critical Rayleigh number, the preferred values are $Ra_c\sim1.8970\times 10^6$ with $k_c\sim28.0243$ for the stress-free case and $Ra_c\sim1.5251\times 10^6$ with $k_c\sim24.6366$ for the no-slip case, for the values of $E$ and $P$ used in figure \ref{fig:Rares}. This is in 
agreement with \citet{cha61}. These critical values of the wavenumbers do however depend on $Re$. In 
the case of $Re=0$ the system has complete symmetry in the $x$ and $y$ directions,
so all wavenumbers $k_x$ and $k_y$ satisfying $k_x^2 + k_y^2 = k_c^2$
onset at $Ra_c$. However as the zonal wind strength is increased from zero 
we found there is immediately a preference for two-dimensional modes with $k_{y_c}=0$. This is the case for all modes with $Re\ne0$. We also find
that the value of the critical Rayleigh number decreases, for both types of boundary conditions, as shown by figure \ref{fig:Racrits}. 
Hence the zonal wind has a destabilising 
effect on the system and aids the onset of convection. The critical azimuthal wavenumber, $k_{x_c}$, also decreases as $Re$ is increased for both types of boundary conditions as shown by figure \ref{fig:Racrits}. The two plots of eigenfunctions in the convective regime, $\times_1$ and $\times_4$ are for critical values of $k_x$ and $Ra^*$ with $Re=5$. 

As $Re$ is increased we move into the baroclinic regime and hence the values of $Re$ chosen for the plot in figure \ref{fig:Racrits} are relatively low in order to remain in the convective regime. For the modes in the convective regime the main energy balance is between 
the buoyancy and the viscous stresses. However as $Re$ is increased, the baroclinic basic state means that
buoyancy can do work at lower critical Rayleigh number, and indeed even at negative Rayleigh number. 
This is discussed in section \ref{sec:energy}.

\begin{figure}[t]
\centering
\hspace{-10pt}\subfigure[Stress-free boundaries.]{\label{fig:Recrit}\includegraphics[width=0.50\textwidth,angle=0]{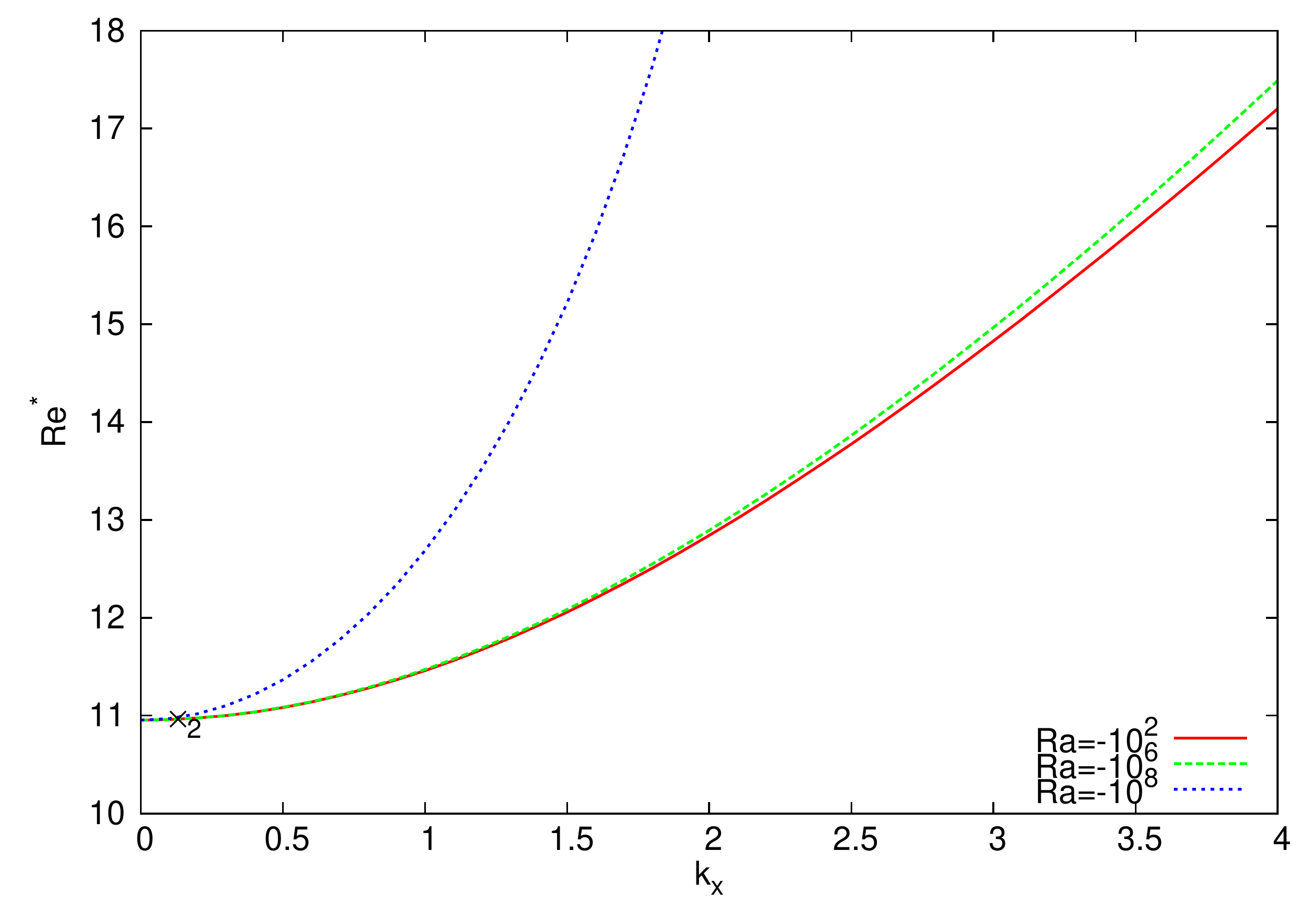}}
\subfigure[No-slip boundaries.]{\label{fig:Recrit2}\includegraphics[width=0.50\textwidth,angle=0]{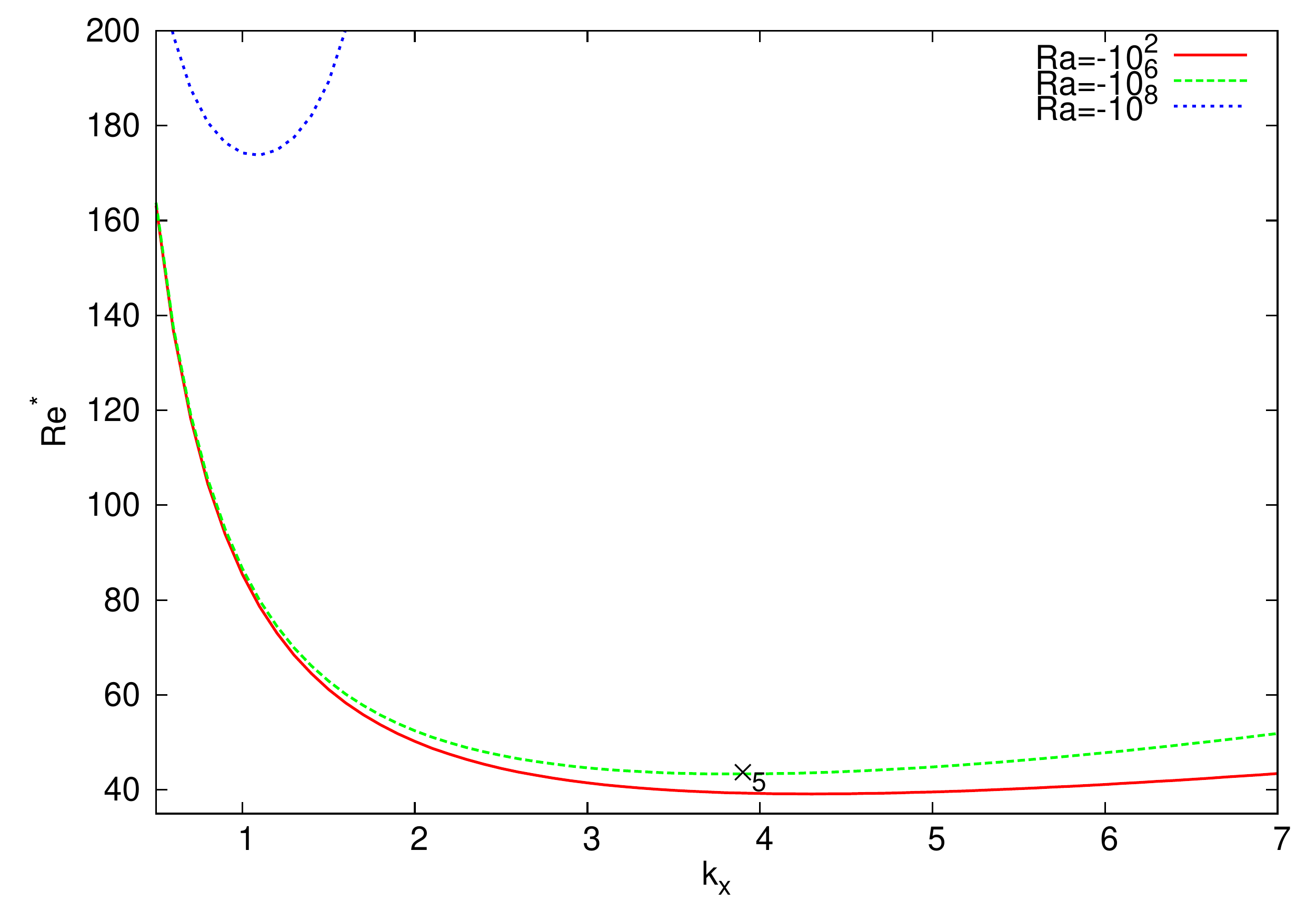}}
\caption{Plots of the numerical results for the onset parameters in the baroclinic regime against $k_x$ with 
$E=10^{-4}$, $P=1$, $k_y=k_{y_c}=0$. The onset parameter is the Reynolds number in the baroclinic regime.}
\label{fig:Recrits}
\end{figure}

\subsection{Baroclinic regime} \label{sec:barreg}

As the zonal wind strength is increased further we find a second type of mode, which is interesting as it 
allows for instability regardless of how negative the Rayleigh number is. In other words this mode can 
be unstable no matter how stably stratified the system is. For this reason we refer to them 
as `baroclinic modes', which are distinct from the convective modes that are usually found as the 
most unstable modes. They are related to the unstable modes of the Eady
problem \citep{ped87}. This suggests that we should consider a critical Reynolds number, rather than a critical Rayleigh number, for the baroclinic modes since it is the shear that is driving this instability. Hence we introduce a critical Reynolds number, $Re_c$, and corresponding critical wavenumbers, $k_{x_c}$ and $k_{y_c}$ for the baroclinic regime. For a given Ekman number, Prandtl number and Rayleigh number $Re_c$ is the value of the Reynolds number for which a marginal baroclinic mode can appear (analogous to the critical Rayleigh number in the convective regime). As with all modes with a non-zero Reynolds number we find that $k_{y_c}=0$. From figure \ref{fig:Recrits} we see how $Re^*$ varies with $k_x$ for several negative values of the Rayleigh number for both types of boundary conditions. 

For stress-free boundaries we see from figure \ref{fig:Recrit} that in all cases $k_{x_c}=0$ and $Re_c\sim10.95$. Therefore reducing $k_x$ allows for instability with an ever more negative Rayleigh number as shown by table \ref{tab:Ek}. It is for this reason that $Ra^*$ rather than $Ra_c$ is plotted in figure \ref{fig:Rares}. An asymptotic theory highlighting these results and which obtains a value of $Re_c$ for any given $Ra$ and $P$ in the small $E$ limit, is discussed in section \ref{sec:Reasymp}. The form of a typical baroclinic mode at onset is shown in figure \ref{fig:x2}, point $\times_2$. We see that the vorticity is independent of $z$ and that $\theta$ has flipped signs for this type of mode so that the hot fluid is sinking and 
the cold fluid is rising. This is directly related to the change in sign of the Rayleigh number and 
is due to the fact that the baroclinic basic state allows buoyancy to fully balance the viscous stresses 
even at negative Rayleigh number (see section \ref{sec:energy}). However the magnitude of the vertical velocity is small, indicating that the shear is dominating the flow in these modes. The form of the eigenfunctions suggest that an asymptotic analysis may be possible for small $k_x$, which is developed in section \ref{sec:asym}. The general form of the eigenfunctions remains similar
to that shown in figure \ref{fig:x2} as $k_x$ is reduced towards the true critical value namely $k_{x_c}=0$.

\begin{table}[b]
\centering
\begin{tabular}{cc|c|c|c|}
\cline{2-4}
&  \multicolumn{3}{|c|}{$Ra^*$} \\
\hline
\multicolumn{1}{|c|}{$k_x$} & $E=10^{-3}$ & $E=10^{-4}$ & $E=10^{-5}$ \\ \hline\hline
\multicolumn{1}{|c|}{0.01} & $-9.6562\times 10^{10}$ & $-9.6578\times 10^{12}$ & $-9.6577\times 10^{14}$ \\
\multicolumn{1}{|c|}{0.05} & $-3.8618\times 10^{9}$ & $-3.5057\times 10^{11}$ & $-3.8624\times 10^{13}$ \\
\multicolumn{1}{|c|}{0.1} & $-9.6496\times 10^{8}$ & $-9.6511\times 10^{10}$ & $-9.6511\times 10^{12}$ \\
\multicolumn{1}{|c|}{0.5} & $-3.7972\times 10^{7}$ & $ -3.7980\times 10^{9}$ & $ -3.7980\times 10^{11}$ \\
\multicolumn{1}{|c|}{1} & $-9.0591\times 10^{6}$ & $-9.0636\times 10^{8}$ & $-9.0637\times 10^{10}$ \\
\hline
\end{tabular}
\vspace{20pt}
\caption{Numerically computed values of $Ra^*$ for various $E$ and $k_x$ in the case $Re=100$, $P=1$ and $k_y=k_{y_c}=0$
for stress-free boundaries.}
\label{tab:Ek}
\end{table}

For no-slip boundaries we see from figure \ref{fig:Recrit2} that there is a non-zero critical azimuthal wavenumber, which varies with $Ra$. As the Rayleigh number is made more negative the critical azimuthal wavelength lengthens and the critical Reynolds number increases. Figure \ref{fig:x5}, point $\times_5$, shows the form of the eigenfunctions at critical for $Ra=-10^6$. As with the stress-free case the sign of $\theta$ has changed from the convective regime and the magnitude of $u_z$ is small. However the vorticity now takes a more complicated slanted structure, which is asymmetric in $z$, in contrast to the stress-free case where $\omega_z$ was independent of $z$.

The baroclinic modes are only found for certain parameter regimes as highlighted by figure \ref{fig:Rares}. For stress-free boundaries we must 
have $k_x \lesssim 30$ and $Re \gtrsim 10$ for these modes to appear and as such this is a constraint on the 
existence of the baroclinic modes. For no-slip boundaries the parameter regime for the existence 
of the baroclinic modes is altered slightly but we still require a sufficiently large $Re$ and 
sufficiently small $k_x$. Outside of these regimes we recover the convective modes, which have 
positive Rayleigh number. This is demonstrated by considering the $Re=1$ line in figure \ref{fig:Raressf}, 
which has solely positive $Ra^*$. In the stress-free case, for a sufficiently large $Re$, the Rayleigh number 
is negative and depends on $k_x$ and $E$ such that reducing either of these parameters towards 
zero makes the Rayleigh number more negative, thus making the system less stable. In fact 
from table \ref{tab:Ek} it is clear that the magnitude of $Ra^*$ is inversely proportional 
to both $k_x^2$ and $E^2$. This remains true for different values of $Re$. In this way we see that 
it is possible to have instability regardless of how negative the Rayleigh number is by choosing a 
small enough $k_x$ and sufficiently large $Re$.

\subsection{Further numeric results}
\label{sec:morenum}

Between the regions of positive and negative Rayleigh number there is a sharp transition region 
where the Rayleigh number passes through zero in a relatively small region of $Re$-space.  
The Rayleigh number varies smoothly from positive to negative values across the transition region.
The values of the Reynolds number at onset, in the case of stress-free boundaries, for a given $k_x$, $Re^*$, for the transition region at which $Ra^*=0$ are given in table \ref{tab:transnum}. As $E$ is reduced $Re^*$ at transition converges to a value independent of the Ekman number. From table \ref{tab:transnum} we also notice that reducing $k_x$ lowers the Reynolds number at onset suggesting once again that the minimising $k_x$ is zero (i.e. $k_{x_c}=0$) and $Re_c$ is converging to a value dependent on the Prandtl number.

The modes described so far have all been steady. Steady modes are usually preferred for the 
onset of convection in a plane layer at $P=1$, unsteady modes being possible at lower
$P$ \citep{cha61}. However by increasing $Re$ further we also found unsteady modes 
appearing at onset even at $P=1$. 
These modes are found in the region of parameter space shown in figures \ref{fig:Raressf}
and  \ref{fig:Raresns} to the right of the dividing curve,
the solid line in both figures. We see that these unsteady modes can onset with either 
positive or negative Rayleigh number. Figure \ref{fig:x3}, point $\times_3$, shows the eigenfunctions 
for such an oscillatory mode in the case of stress-free boundaries. These modes onset as pairs of travelling 
wall modes with frequencies which are equal but opposite in sign. Oscillatory modes are found at
larger $k_x$ and $Re$ for the no-slip case, an example being shown in figure \ref{fig:x6}, point $\times_6$.
If the domain is infinite in the $x$ and $y$ directions, all wavenumbers $k_x$ and $k_y$ are allowed,
and the critical mode is always steady, either at fixed $Ra$ as $Re$ is gradually increased or at
fixed $Re$ as $Ra$ is gradually increased. However, if the domain is finite, and for example
periodic boundary conditions in $x$ and $y$ are imposed, thus restricting the possible choice of
wavenumbers to a discrete set, then it is possible for oscillatory modes to be preferred.

\begin{table}[t]
\centering
\begin{tabular}{ccccccccc}
\cline{2-9}
& \multicolumn{8}{|c|}{$Re^*$} \\
\cline{2-9}
& \multicolumn{4}{|c|}{$E=10^{-4}$} &
\multicolumn{4}{|c|}{$E=10^{-5}$} \\
\hline
\multicolumn{1}{c}{$k_x$} & $P=0.1$ & $P=1$ & $P=10$ & $P=20$ & $P=0.1$ &
$P=1$ & $P=10$ & $P=20$ \\ \hline\hline
\multicolumn{1}{c}{0.1} & 34.871848	& 10.961025 &	3.464601 & 
1.538575 &	34.694565 &	10.955008 &	3.464401 &	1.538771 \\
\multicolumn{1}{c}{0.5} & 35.028908 & 11.073052 & 3.470478 & 
1.505469 & 34.932575 & 11.025471 & 3.468142 & 1.510741 \\
\multicolumn{1}{c}{1.0} & 36.362040 & 11.626575 & 3.526367 & 
1.428312 & 36.309677 & 11.612943 & 3.520017 & 1.426985 \\
\multicolumn{1}{c}{5.0} & 64.790420 & 19.831849 & 5.088978 & 
1.591699 & 64.777473 & 19.829615 & 5.088115 & 1.591423 \\
\multicolumn{1}{c}{10.0} & 115.463528 & 35.190378 & 10.557187 & 
4.845180 & 114.698854 & 35.120826 & 10.512839 & 4.806599 \\
\hline
\end{tabular}
\caption{Numeric results showing the position of the transition region,
the point where $Ra^*=0$,
in $Re$-space for various values of $k_x$, $E$ and $P$ in the case
$k_y=k_{y_c}=0$ for stress-free boundaries.}
\label{tab:transnum}
\end{table}


In the work displayed so far we have varied the parameters of most interest: $k_x$, $Re$ and $Ra$ whilst 
looking at specific values for $P$ and $E$. We have also found that $k_{y_c}=0$ for the modes of interest (i.e. modes with $Re\ne0$). Although instability is possible with $k_y\ne0$ in both the convective and baroclinic regimes, we find that increasing $k_y$ from zero only serves to stabilise the system by increasing the Rayleigh number or Reynolds number for which onset occurs. Here we consider the effects of varying the Ekman and Prandtl numbers.

We first look at two further values for the Ekman number: $10^{-3}$ and $10^{-5}$. We find that 
changing $E$ alters the magnitude of the Rayleigh number at onset but does not affect the position 
of the baroclinic parameter regime in $k_x\,-\, Re$ space. The results in table \ref{tab:Ek} highlight the fact that 
for the baroclinic mode $Ra^*$ is inversely proportional to $E^2$. Therefore if we increase the 
Rayleigh number from $-\infty$ changing the Ekman number controls how soon the instability occurs. 
However we still require the same sufficiently large $Re$ and small values of $k_x$.

\begin{figure}[t]
\centering
\includegraphics[width=0.5\textwidth]{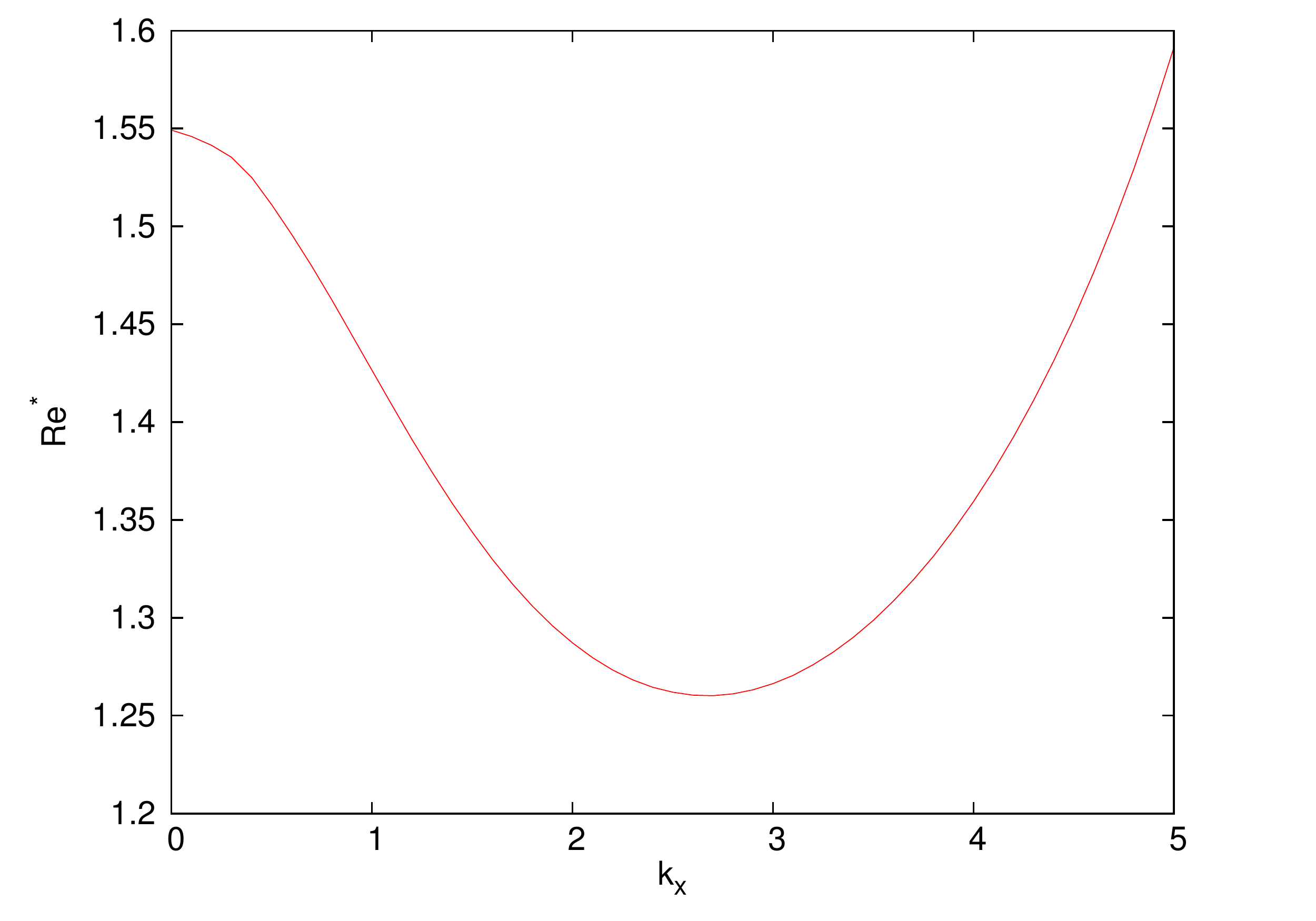}
\caption{Plot showing how the Reynolds number at onset varies with $k_x$ for $P=50$, $E=10^{-4}$, $Ra=-1$ and $k_y=k_{y_c}=0$ with stress-free boundaries.}
\label{fig:P=50plot}
\end{figure}

We considered further values of the Prandtl number: $P=0.1$ and $P=10$. In a way the effect of 
changing the Prandtl number was opposite to that of altering the Ekman number. This is because although 
the Rayleigh number remains largely unaffected for various $P$, the position of the baroclinic regime in $k_x\,-\, Re$ space changes. This can be seen in table \ref{tab:transnum} where the transition region occurs at a higher/lower value of $Re^*$ for a lower/higher value of $P$. We see that for $P=10$ the 
baroclinic modes are able to appear at a lower value of the zonal wind ($Re\sim 3.5$), compared to 
the $P=1$ case. The converse is true when $P=0.1$ where the baroclinic modes cannot appear until $Re\sim 35$. The behaviour of the critical parameters at moderate values of the Prandtl number $(P=0.1-10)$ remains largely the same with $k_{x_c}=0$ continuing to be preferred in the stress-free baroclinic regime. However we note that there is a non-zero miminising $k_x$ for larger values of $P$ so long as the magnitude of $Ra$ is not too large. An example of this can be seen in table \ref{tab:transnum} when $P=20$, for both values of the Ekman number. Another case, with $Ra$ non-zero, is displayed in figure \ref{fig:P=50plot} where we find $k_{x_c}\sim2.7$ for $P=50$ with $Ra=-1$. The critical value of the Reynolds number is $\sim1.2603$, which is smaller than for the other Prandtl numbers considered, as expected. The asymptotic theory in section \ref{sec:Reasymp} is able to explain this dependence of $k_{x_c}$ on $P$.


\subsection{Thermodynamic equation} \label{sec:energy}

\begin{figure}[t]
\centering
\includegraphics[width=0.5\textwidth]{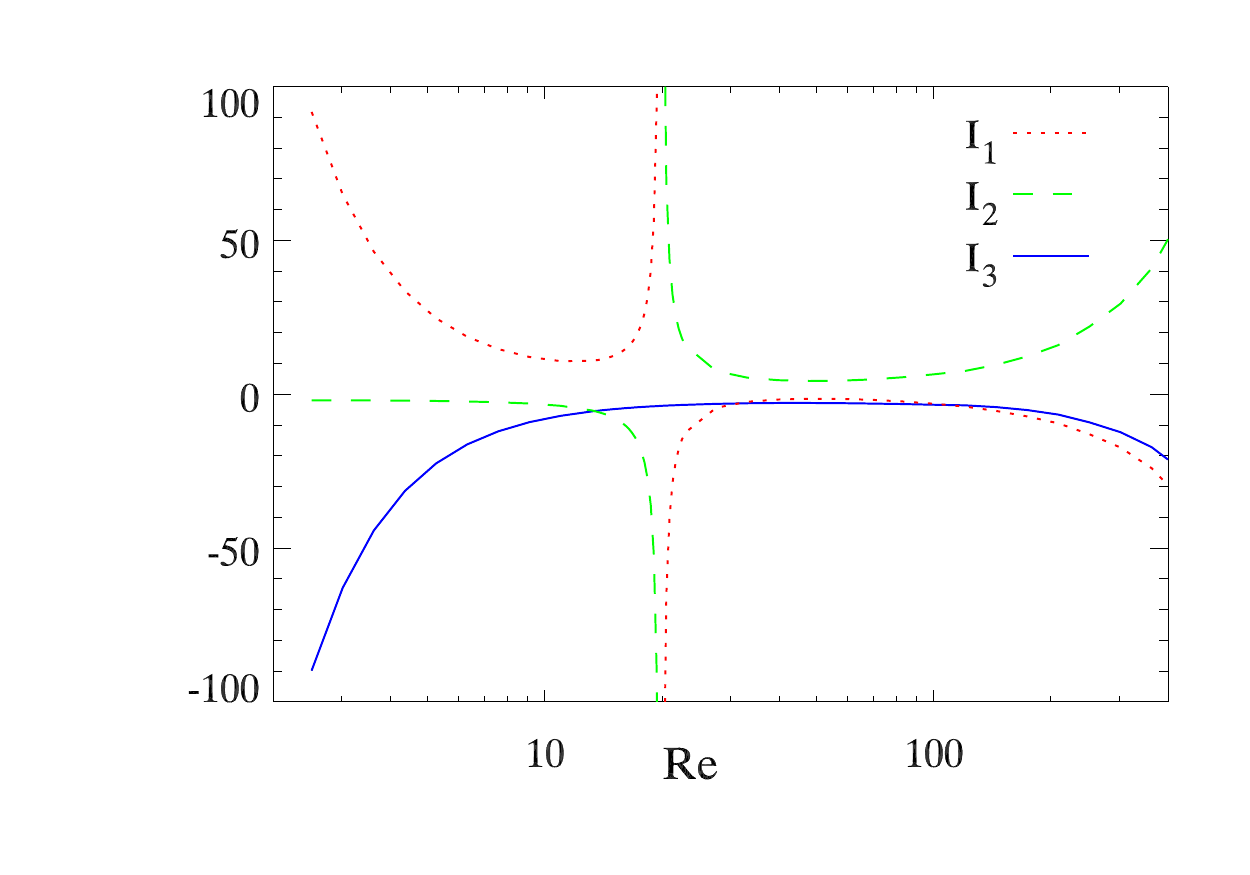}
\caption{Plot showing how the integrals in the thermodynamic equation  (\ref{eq:thermo}) vary
 with $Re$, at $E=10^{-4}$. There are stress-free boundaries 
and $P=1$, $k_x=5.0119$ and $k_y=k_{y_c}=0$.}
\label{fig:thermo}
\end{figure}

To form the energy equation we consider the dot product of $\mathbf{u}$ with equation (\ref{eq:nsptb2}) and 
integrate over the volume of the layer. In the limit $k_y=0$, the situation
most favourable to baroclinic instability, the only terms that remain are the balance of the work
done by buoyancy and the rate of working of the viscous forces,
\begin{equation}
Ra \int u_z \theta \,dV = \int\frac{\partial u_i}{\partial x_j}\frac{\partial u_i}{\partial x_j}\,dV,
\label{eq:emom}
\end{equation}
where we have non-dimensionalised using the same scales as earlier. Following chapter 2 of
\cite{cha61}, we now multiply the temperature equation (\ref{eq:tempptb2}) by $\theta$ 
and eliminate the rate of working of the buoyancy force to obtain the thermodynamic equation
\begin{equation}
Ra\int(\nabla\theta)^2dV - PE^{-1}Re\int\theta u_y dV - \int\frac{\partial u_i}{\partial x_j}
\frac{\partial u_i}{\partial x_j}dV = 0, \quad {\rm or} \quad I_1 + I_2 + I_3 =0.
\label{eq:thermo}
\end{equation}
The second integral, $I_2$, is related to the heat flux carried in the $y$ direction, and is only non-zero
when the zonal flow is non-zero. The third term, $I_3$, is the rate of  
viscous dissipation. We can write 
this equation in terms of the real and imaginary parts of $u_z$, $\omega_z$ and $\theta$ and 
their derivatives, all of which have been calculated in the numerics above. Figure \ref{fig:thermo} 
shows how the three terms in equation (\ref{eq:thermo}) vary as a function of $Re$ for a specific 
choice of $k_x$. Plots for other $k_x$ where baroclinic modes exist are similar with the position 
of the transition region changing accordingly.

Since the integral in the first term is positive definite, and that in the third term
is negative definite, we must also have $Ra>0$ 
in the case $Re=0$. This is the well understood case where the Rayleigh number must be positive for 
the system to be convectively unstable. At low $Re$ this remains the predominant balance and the 
Rayleigh number remains positive. However with $Re\neq0$ the baroclinic term can now 
partially balance the viscous stresses and thus as $Re$ is increased the Rayleigh number is 
reduced to allow equation (\ref{eq:thermo}) to balance. This can be seen in figure \ref{fig:thermo} 
where the $I_2$ contribution slowly increases in magnitude as $Re$ increases.

As $Re$ is increased further and we enter the transition region (located at $Re\sim 19.86$ 
for $k_x=5.0119$) we see that both $I_1$ and the baroclinic flux, $I_2$, change sign. In the transition region the main balance is between these two terms as the magnitude of the rate of working of the viscous stresses is small. However the sum of $I_1$ and $I_2$ must still balance the always negative $I_3$ term. The transition region represents the point in $Re$-space where $I_2$ becomes large enough in magnitude 
to solely overcome $I_3$ without the need for a contribution from $I_1$. Hence $I_1$ can change sign, so that  
$Ra$ must change sign also. This explains why a sufficiently large value of the zonal wind is required 
to allow for modes with negative Rayleigh number to appear. It also indicates that the term, $I_1$ or $I_2$, in equation (\ref{eq:thermo}) which is positive, and thus is able to balance $I_3$, contains the parameter that is driving the instability. In other words it is the Rayleigh/Reynolds number and thus the work done by buoyancy/baroclinic heat flux, which is balancing the 
viscous 
dissipation in the convective/baroclinic regime.

Equation (\ref{eq:thermo}) can also explain the results of changing the Prandtl number given by 
table \ref{tab:transnum}. Since $I_2$ in the thermodynamic equation is proportional to $P$, 
increasing or decreasing the Prandtl number requires a lower or higher value of $Re$ respectively. 
This is slightly crude since it assumes that the values of the integrals in equation (\ref{eq:thermo}) 
do not change with $P$. This is not the case, which is why increasing the Prandtl number by an order 
of magnitude does not result in the zonal wind decreasing by the same amount. For example the position 
of the transition region for $P=10$ in table \ref{tab:transnum} has only moved from $Re\sim10$ (in the $P=1$
case) to $Re\sim3.5$ 
rather than $Re\sim1$. Despite this the form of $I_2$ in the thermodynamic equation 
serves to explain the general dependency of the transition region on $P$.

\section{Asymptotics} \label{sec:asym}

Here we develop asymptotic theories, which predict the numeric results with stress-free boundaries very well. In the numerical work previously discussed we have been considering low but finite values of the Ekman number since these are of particular physical interest. Hence the first limit to take is that of asymptotically small $E$. Guided by the numerics we rescale the dependent variables as
$\omega_z = {\tilde \omega_z} $, $u_z= E {\tilde u_z}$, $\theta = E {\tilde \theta}$, and $Ra = -{\tilde R}/ E^2$ and then we find that the leading order equations from (\ref{eq:vortbc}) - (\ref{eq:tempbc}) in the limit $E\to 0$ are
\begin{gather}
\left(\sigma + \mathrm{i}k_x Re z + k^2 -\frac{\mathrm{d}^2}{\mathrm{d}z^2}\right){\tilde \omega_z}  = \frac{\mathrm{d}{\tilde u_z}}{\mathrm{d}z}, \label{eq:asymEvort}\\
\frac{\mathrm{d}{\tilde \omega_z}}{\mathrm{d}z} = 
k^2{\tilde R}{\tilde  \theta}, \label{eq:asymEcurlvort}\\
\left(\sigma + \mathrm{i}k_x P Re z + k^2 - \frac{\mathrm{d}^2}{\mathrm{d}z^2}\right){\tilde \theta} = {\tilde u_z}
 + \frac{\mathrm{i}k_x P Re}{k^2{\tilde R}} {\tilde \omega_z}. \label{eq:asymEtemp}
\end{gather}
From these equations we are able to easily eliminate $\tilde{\theta}$ by taking the double-derivative of equation  
(\ref{eq:asymEcurlvort}) and substituting into equation (\ref{eq:asymEtemp}) to give
\begin{equation}
\frac{\mathrm{d}^3{\tilde \omega_z}}{\mathrm{d}z^3} = (\sigma + \mathrm{i}k_xPRe z + k^2)\frac{\mathrm{d}{\tilde \omega_z}}{\mathrm{d}z} - \mathrm{i}k_xPRe{\tilde \omega_z} - \tilde{R}k^2\tilde{u}_z.
\label{eq:asymE3ord}
\end{equation}
We use equations (\ref{eq:asymEvort}) - (\ref{eq:asymE3ord}) in each of our asymptotic theories and thus they are only accurate for low Ekman numbers. These equations are related to the quasi-geostrophic equations used by atmospheric scientists, see section 4.3 below, though here diffusion is still included.

\subsection{Low wavenumber asymptotics 1: Fixed Rayleigh number}
\label{sec:Reasymp}

In this theory we obtain an expression for the critical Reynolds number in terms of the Prandtl and Rayleigh numbers. 
We set $\sigma =0$ because we are considering steady marginal modes and since the critical latitudinal wavenumber vanishes for all modes of interest we also set $k_y=k_{y_c}=0$ so that $k=k_x$. The numerics suggest that the critical azimuthal wavenumber is zero for baroclinic modes and thus we expand $\tilde{\omega}_z$, $\tilde{u}_z$ and $Re$ in powers of the small parameter $k$ as follows:
\begin{eqnarray}
{\tilde \omega_z} &=& 1 + k\omega_1 + k^2\omega_2 + \cdots , \\
{\tilde u_z} &=& k(u_0 + k u_1 + k^2 u_2 + \cdots) , \\
Re &=&  Re_0 + k Re_1 + k^2 Re_2 + \cdots. \label{eq:Reexp}
\end{eqnarray}
We have chosen $\omega_0=1$ to satisfy normalisation conditions and in this theory we apply the stress-free boundary conditions given by equation (\ref{eq:sfbcs}). It is also useful to take the integral of equation (\ref{eq:asymEvort}) across the layer since the boundary conditions eliminate two of the resulting terms to leave
\begin{equation}
\int_{-1/2}^{1/2} (\mathrm{i}kRez+k^2)\tilde{\omega}_z \mathrm{d}z= 0.
\label{eq:asymint}
\end{equation}
We now proceed by considering the equations at increasing order (i.e. powers of $k$). Equations (\ref{eq:asymE3ord}) and (\ref{eq:asymEvort}) at order $\mathrm{O}(k)$ give
\begin{gather}
\frac{\mathrm{d}^3\omega_1}{\mathrm{d}z^3} = -\mathrm{i}PRe_0, \\
\mathrm{i}Re_0z-\frac{\mathrm{d}^2\omega_1}{\mathrm{d}z^2} = \frac{\mathrm{d}u_0}{\mathrm{d}z}
\end{gather}
respectively, which when applying the boundary conditions gives
\begin{gather}
\omega_1 = -\mathrm{i}PRe_0\left(\frac{z^3}{6}-\frac{z}{8}\right), \label{eq:om1} \\
u_0 = \mathrm{i}Re_0(1+P)\left(\frac{z^2}{2}-\frac{1}{8}\right).
\end{gather}
Next we consider equations (\ref{eq:asymint}) and (\ref{eq:asymE3ord}) at $\mathrm{O}(k^2)$ to give
\begin{gather}
\int_{-1/2}^{1/2} (\mathrm{i}Re_0z\omega_1 + \mathrm{i}Re_1z + 1) \mathrm{d}z = 0, \label{eq:asymEintk2} \\
\frac{\mathrm{d}^3\omega_2}{\mathrm{d}z^3} = \mathrm{i}PRe_0z\frac{\mathrm{d}\omega_1}{\mathrm{d}z} - \mathrm{i}PRe_0\omega_1 -\mathrm{i}PRe_1  \label{eq:asymE3ordk2}
\end{gather}
and by using the definition of $\omega_1$ we can evaluate the integral in equation (\ref{eq:asymEintk2}) to acquire
\begin{equation}
Re_0 = \sqrt{\frac{120}{P}}.
\label{eq:Re0}
\end{equation}
We can also find $\omega_2$ from equation (\ref{eq:asymE3ordk2}) by inserting the definition of $\omega_1$ and using the boundary conditions to get
\begin{equation}
\omega_2 = P^2Re_0^2\left(\frac{z^6}{360} - \frac{z^2}{1920}\right) - \mathrm{i}PRe_1\left(\frac{z^3}{6} - \frac{z}{8}\right).
\label{eq:om2}
\end{equation}
Once again considering equations (\ref{eq:asymint}) and (\ref{eq:asymE3ord}), now at $\mathrm{O}(k^3)$, we obtain
\begin{gather}
\int_{-1/2}^{1/2} (\mathrm{i}Re_0z\omega_2 + \mathrm{i}Re_1z\omega_1 + \mathrm{i}Re_2z + \omega_1) \mathrm{d}z = 0, \label{eq:asymEintk3} \\
\frac{\mathrm{d}^3\omega_3}{\mathrm{d}z^3} = \mathrm{i}PRe_0z\frac{\mathrm{d}\omega_2}{\mathrm{d}z} + \mathrm{i}PRe_1z\frac{\mathrm{d}\omega_1}{\mathrm{d}z} +\frac{\mathrm{d}\omega_1}{\mathrm{d}z} - \mathrm{i}PRe_0\omega_2 -\mathrm{i}PRe_1\omega_1 -\mathrm{i}PRe_2 - \tilde{R}u_0 \label{eq:asymE3ordk3}
\end{gather}
respectively whereby $Re_1=0$ to satisfy equation (\ref{eq:asymEintk3}). By inserting the definitions of $\omega_1$, $\omega_2$ and $u_0$ into equation (\ref{eq:asymE3ordk3}) we find
\begin{equation}
\omega_3 = \mathrm{i}P^3Re_0^3\left(\frac{z^9}{36288} - \frac{z^5}{115200} + \frac{z}{573440}\right) - \mathrm{i}Re_0(P + \tilde{R}(1+P))\left(\frac{z^5}{120} - \frac{z^3}{48} + \frac{5z}{384}\right) - \mathrm{i}PRe_2\left(\frac{z^3}{6} - \frac{z}{8}\right).
\label{eq:om3}
\end{equation}
We are now able to find an expression for $Re_2$ using equation (\ref{eq:asymint}) at $\mathrm{O}(k^4)$, which is
\begin{equation}
\int_{-1/2}^{1/2} (\mathrm{i}Re_0z\omega_3 + \mathrm{i}Re_2z\omega_1 + \mathrm{i}Re_3z + \omega_2) \mathrm{d}z = 0. \label{eq:asymEintk4}
\end{equation}
We insert the expressions for $\omega_1$, $\omega_2$ and $\omega_3$ into equation (\ref{eq:asymEintk4}) and evaluate the integral to find
\begin{equation}
Re_2 = \sqrt{\frac{30}{P}}\left[\frac{17}{168}\left(1 + \frac{\tilde{R}(1+P)}{P}\right) - \frac{5P}{792}\right].
\label{eq:Re2}
\end{equation}
Hence from equation (\ref{eq:Reexp}) we find
\begin{equation}
Re \approx Re_0 + k^2Re_2 = \sqrt{\frac{120}{P}} + k^2\sqrt{\frac{30}{P}}\left[\frac{17}{168}\left(1 + \frac{\tilde{R}(1+P)}{P}\right) - \frac{5P}{792}\right],
\label{eq:Reexpress}
\end{equation}
which yields an approximation to the Reynolds number given $P$, $\tilde{R}$ and a small $k$. The form of this expression for $Re$ is able to explain the dependence of the critical wavenumber on $P$ as seen in section \ref{sec:morenum}. For a given Prandtl number the $Re_0$ term in the expression for $Re$ given by (\ref{eq:Reexpress}) gives an approximation to the critical Reynolds number. For example with $P=1$ this term is $\sim10.9545$, which is in excellent agreement with the numerics discussed in section \ref{sec:barreg}. The second term of equation (\ref{eq:Reexpress}) then gives an adjustment to the the leading order value for $Re$. The sign of this term determines whether $k_{c}=0$ or not. If, for a given $P$ and $\tilde{R}$, the value of $Re_2$ is positive then the adjustment to $Re_0$ can only serve to increase the Reynolds number and hence the preferred value of $k$ to minimise $Re$ is $k=0$ as expected given the numeric results from section \ref{sec:barreg}. However if the value of $Re_2$ is negative (again for given $P$ and $\tilde{R}$) a non-zero $k$ must be preferred as the inclusion of this term now lowers the Reynolds number from the $Re_0$ value.

Table 3 displays quantities for $Re_0$ and $Re_2$ for various values of $P$ and $\tilde{R}$. Since $Re_0$ is independent of $\tilde{R}$ this only varies with $P$ and the values predicted for the Reynolds number match the numerics of table \ref{tab:transnum} very well. For most combinations of $P$ and $\tilde{R}$ the value of $Re_2$ is positive, confirming that $k_c=0$ and $Re_c=Re_0$. However for certain choices of the parameters we obtain negative values for $Re_2$ indicating that there is a non-zero minimising value of $k$. This was seen in the numerics where we recall from figure \ref{fig:P=50plot} that there was a non-zero $k_c$ for $P=50$ and $Ra=-1$. The equivalent values of the Prandtl and Rayleigh numbers in the asymptotic theory ($P=50$ and $\tilde{R}=1$) give a negative value of $Re_2$ agreeing with the numerics that there is a non-zero minimising $k$.

\begin{table}[b]
\centering
\begin{tabular}{cccccc}
\cline{3-6}
& & \multicolumn{4}{|c|}{$Re_2$} \\
\hline
\multicolumn{1}{c}{$P$} & $Re_0$ & $\tilde{R}=0$ & $\tilde{R}=1$ & $\tilde{R}=10$ & $\tilde{R}=1000$ \\ \hline\hline
\multicolumn{1}{c}{0.1} & $34.64102$ & $1.74174$ & $21.02111$ & $194.53549$ & $19281.11679$ \\
\multicolumn{1}{c}{1} & $10.95445$ & $0.51966$ & $1.62815$ & $11.60453$ & $1109.00579$ \\
\multicolumn{1}{c}{10} & $3.46410$ & $0.065920$ & $0.25871$ & $1.99386$ & $192.85967$ \\
\multicolumn{1}{c}{50} & $1.54919$ & $-0.16612$ & $-0.086175$ & $0.63337$ & $79.78332$ \\
\multicolumn{1}{c}{100} & $1.09545$ & $-0.29036$ & $-0.23438$ & $0.26943$ & $55.68819$ \\
 \hline
\end{tabular}
\label{tab:Repredict}
\caption{Table displaying values for $Re_0$ and $Re_2$ for various Prandtl and Rayleigh numbers as given by the 
expression in equation (\ref{eq:Reexpress}).}

\end{table}

This theory is unable to predict the critical wavenumber and critical Reynolds number when $Re_2<0$ without including higher order terms, which would give an $\mathrm{O}(k^4)$ term in equation (\ref{eq:Reexpress}). However it does indicate which values of the Prandtl and Rayleigh numbers we would expect to find a non-zero critical wavenumber for and it predicts $Re_c$ very accurately for the $k_c=0$ cases.


We are also able to solve equations (\ref{eq:asymEvort}) and (\ref{eq:asymE3ord}) numerically without the assumption of small $k$. Using stress-free boundary conditions and the normalisation and symmetry conditions of the eigenfunctions (known from the numerics) we have a fourth order complex BVP with nine real boundary conditions, including a normalisation condition,
\begin{gather}
\omega_\mathrm{r}(0)=1,\ \ \omega_\mathrm{i}(0)=\omega_\mathrm{r}^\prime(0)=\omega_\mathrm{i}^{\prime\prime}(0)=u_\mathrm{r}(0)=0, \ \
\omega_\mathrm{r}^\prime(0.5)=\omega_\mathrm{i}^\prime(0.5)=u_\mathrm{r}(0.5)=u_\mathrm{i}(0.5)=0.
\label{eq:transbcs}
\end{gather}
Here the primes and subscripts indicate the derivatives and the real and imaginary parts of the 
eigenfunctions respectively. The system defined by (\ref{eq:asymEvort}), 
(\ref{eq:asymE3ord}) and (\ref{eq:transbcs}) 
is an eighth order homogeneous system in the real variables, with eight homogeneous boundary
conditions and a normalisation condition, so it has an eigenvalue, $Re$. Hence given specific values of $k$, $P$ and $\tilde{R}$ we can find a value for $Re$.
We solved this system using a simple BVP solver in Maple and some results for the case of $\tilde{R}=0$ are displayed in table 4(a). By comparing the values of $Re$ in table 4(a) with the location of the transition region from figure \ref{fig:Raressf} and table \ref{tab:transnum} we see that the asymptotic theory predicts the location 
of the transition region very well. In particular, we see that the position of the transition region is converging,
as we reduce $E$, to a value similar to that predicted by the asymptotics in all cases. Also of note is that for $P=20$ and $P=50$ there are minimising values of the azimuthal wavenumber due to the fact that the Rayleigh number is small enough to allow this to occur. In fact this can be checked by evaluating $Re_2$ as given by equation (\ref{eq:Reexpress}) with $\tilde{R}=0$ where we indeed find that $Re_2<0$ for $P=20$ and $P=50$, indicating a non-zero critical wavenumber is preferred.

\begin{table}[b]
\label{tab:asym}
\centering
\subtable[]{\begin{tabular}{cccccc}
\cline{2-6}
&  \multicolumn{5}{|c|}{$Re$} \\
\hline
\multicolumn{1}{c}{$k_x$} & $P=0.1$ & $P=1$ & $P=10$ & $P=20$ & $P=50$ \\ \hline\hline
\multicolumn{1}{c}{0.01} & $34.64108$ & $10.95447$ & $3.46410$ & $2.44948$ & $1.54917$ \\
\multicolumn{1}{c}{0.10} & $34.65831$ & $10.95961$ & $3.46475$ & $2.44918$ & $1.54754$ \\
\multicolumn{1}{c}{0.50} & $35.07369$ & $11.08347$ & $3.48056$ & $2.44245$ & $1.51100$ \\
\multicolumn{1}{c}{1.00} & $36.34019$ & $11.46042$ & $3.52963$ & $2.42794$ & $1.42646$ \\
\multicolumn{1}{c}{1.50} & $38.35637$ & $12.05843$ & $3.61070$ & $2.42013$ & $1.34371$ \\
\multicolumn{1}{c}{2.00} & $41.00963$ & $12.84220$ & $3.72316$ & $2.43154$ & $1.28719$ \\
\multicolumn{1}{c}{2.50} & $44.18317$ & $13.77610$ & $3.86686$ & $2.46939$ & $1.26193$ \\
\multicolumn{1}{c}{3.00} & $47.77158$ & $14.82920$ & $4.04217$ & $2.53699$ & $1.26634$ \\
\multicolumn{1}{c}{5.00} & $64.77836$ & $19.83002$ & $5.08823$ & $3.14797$ & $1.59143$ \\
\multicolumn{1}{c}{10.0} & $114.69132$ & $35.11997$ & $10.51240$ & $7.57080$ & $4.80621$ \\
\hline
\label{tab:trans}
\end{tabular}}
\hspace{40pt}
\subtable[]{
\begin{tabular}{ccc}
\hline
$Re$& $k_y/k_x$ & $R_0$ \\ \hline\hline
\multirow{3}{*}{10} & 0 & 0.8983 \\
& 0.001 & 0.8987 \\
& 0.01 & 0.9377 \\ \hline
\multirow{3}{*}{100} & 0 & -9.6578 \\
& 0.001 & -9.5599 \\
& 0.01 & -4.7143 \\ \hline
\multirow{3}{*}{1000} & 0 & -9.8903 \\
& 0.001 & -4.9440 \\
& 0.01 & -0.09560 \\ \hline
\multirow{3}{*}{10000} & 0 & -9.8926 \\
& 0.001 & -0.09792 \\
& 0.01 & -9.6568$\times 10^{-4}$ \\ \hline
\label{tab:lwares}
\end{tabular}}
\caption{Tables displaying results from the two asymptotic theories 
described in section \ref{sec:asym}. 4(a): 
Values for the Reynolds number for various $k_x$ and $P$ in the case $\tilde{R}=0$ found by solving 
the BVP described by equations (\ref{eq:asymEvort}) and (\ref{eq:asymE3ord}). 4(b): Values for $R_0$ found 
by solving equation (\ref{eq:R0solver}) for various values of $Re$ and $k_y/k_x$ with $P=1$.}
\end{table}

\subsection{Low wavenumber asymptotics 2: Fixed Reynolds number}

Here we shall develop an asymptotic theory for the onset of instability at low $k_x$ with stress-free 
boundaries, which predicts the large negative Rayleigh numbers and eigenfunctions very well. Our starting point is 
equations (\ref{eq:asymEvort}) - (\ref{eq:asymEtemp}) with the boundary 
conditions given by (\ref{eq:sfbcs}). We set $\sigma=0$ for the same reason as in section \ref{sec:Reasymp}. The numerics inform us that the baroclinic modes exist for small $k_xRe$ and that we should rescale the Rayleigh number as $\tilde{R} = -{\hat R}/k_x^2$. The equations (\ref{eq:asymEvort}) - (\ref{eq:asymEtemp}) become
  
\begin{gather}
\left(\mathrm{i}k_x Re z + k^2 -\frac{\mathrm{d}^2}{\mathrm{d}z^2}\right){\tilde \omega_z}  - \frac{\mathrm{d}{\tilde u_z}}{\mathrm{d}z} = 0, \label{eq:vortbc2}\\
\frac{\mathrm{d}{\tilde \omega_z}}{\mathrm{d}z} = 
- {\hat R}{\tilde  \theta}, \label{eq:curlvortbc2}\\
\left(\mathrm{i}k_x P Re z + k^2 - \frac{\mathrm{d}^2}{\mathrm{d}z^2}\right){\tilde \theta} = {\tilde u_z}
 - \frac{\mathrm{i}k_x P Re}{{\hat R}} {\tilde \omega_z}. \label{eq:tempbc2}
\end{gather}
We introduce a small parameter $\epsilon$, measuring the magnitude of the horizontal
wavenumbers, and let
\begin{eqnarray}
{\tilde \omega_z} &=& \omega_z^{(0)} + \epsilon \omega_z^{(1)} + \epsilon^2\omega_z^{(2)} + \cdots , \\
{\tilde u_z} &=& \epsilon (u_z^{(0)} + \epsilon u_z^{(1)} + \epsilon^2 u_z^{(2)} + \cdots), \\
{\tilde \theta} &=& \epsilon (\theta^{(0)} + \epsilon \theta^{(1)} + \epsilon^2\theta^{(2)} + \cdots), \\
{\hat R} &=&  R_0 + \epsilon R_1 + \epsilon^2 R_2 + \cdots , \\
k_x &=& \epsilon {\tilde k_x}, \qquad k_y = \epsilon {\tilde k_y}, \label{eq:kyexp}
\end{eqnarray}
where we assume that $R_0<0$ since we are considering the stably stratified modes in this asymptotic expansion. 

We insert these expansions into equations (\ref{eq:vortbc2}) - (\ref{eq:tempbc2}) 
and consider the resulting equations in powers of $\epsilon$ since $k_x \ll 1$. 
Hence we first take each equation at $\mathrm{O}(1)$, which yields
\begin{align}
\omega_z^{(0)} = 1. \label{eq:omznorm}
\end{align}
This choice of $\omega_z^{(0)}$ satisfies this set of equations and is chosen to be unity 
to satisfy normalisation conditions. Also note that this form 
for $\omega_z^{(0)}$ satisfies the stress-free boundary conditions on $\omega_z$ given by (\ref{eq:sfbcs}),
and so no thin boundary layer to match these conditions is required.

Next we consider the first order equations, which are equations (\ref{eq:vortbc2}) - (\ref{eq:tempbc2}) at $\mathrm{O}(\epsilon)$ and we find:
\begin{gather}
\mathrm{i}{\tilde k_x} Re z\omega_z^{(0)} -\frac{\mathrm{d}^2\omega_z^{(1)}}{\mathrm{d}z^2} = \frac{\mathrm{d}u_z^{(0)}}{\mathrm{d}z}, \label{eq:vortford}\\
\frac{\mathrm{d}\omega_z^{(1)}}{\mathrm{d}z} = - R_0 \theta^{(0)}, \label{eq:curlvortford}\\
-\frac{\mathrm{d}^2\theta^{(0)}}{\mathrm{d}z^2} = u_z^{(0)} - \frac{\mathrm{i}{\tilde k_x} Re P\omega_z^{(0)}}{R_0}. \label{eq:tempford}
\end{gather}
We integrate (\ref{eq:vortford}) and apply the no penetration and zero temperature boundary conditions,
 using (\ref{eq:curlvortbc2}), 
to
obtain the constant of integration, and insert the expression for $u_z^{(0)}$ into
(\ref{eq:tempford}) to obtain
\begin{align}
\frac{\mathrm{d}^2\theta^{(0)}}{\mathrm{d}z^2} + R_0 \theta^{(0)} = -\frac{\mathrm{i}{\tilde k_x}Re z^2}{2} + \frac{\mathrm{i}{\tilde k_x}Re}{8} 
+ \frac{\mathrm{i}{\tilde k_x} Re P}{R_0}.
\end{align}
The solution to this inhomogeneous second order ODE in $\theta^{(0)}$ is
\begin{equation}
\theta^{(0)} = A\sinh\left(\sqrt{-R_0}z\right)+B\cosh\left(\sqrt{-R_0} z\right) - \frac{\mathrm{i}{\tilde k_x}Re z^2}{2R_0} + \gamma .
\end{equation}
Due to the symmetry of the boundary conditions $A=0$ so in fact
\begin{eqnarray}
\theta^{(0)} &=& B\cosh\left(\sqrt{-{R}_0}z\right) - \frac{\mathrm{i}{\tilde k_x}Re z^2}{2{R}_0} + \gamma, 
\label{eq:the0} \\
u_z^{(0)} &=& B{R}_0\cosh\left(\sqrt{-{R}_0}z\right) + \frac{\mathrm{i}{\tilde k_x}Re}{R_0}(1+P), \label{eq:uz0} \\
\omega_z^{(1)} &=& B\sqrt{-{R}_0}\sinh\left(\sqrt{-{R}_0}z\right) 
+ \frac{\mathrm{i}{\tilde k_x}Re z^3}{6} - {R}_0 \gamma z, \label{eq:omz1}
\end{eqnarray}
where the expressions for $u_z^{(0)}$ and $\omega_z^{(1)}$ have been found via 
equations (\ref{eq:tempford}) and (\ref{eq:curlvortford}) respectively. 
Any constant of integration in (\ref{eq:omz1}) can be absorbed in the
normalisation condition (\ref{eq:omznorm}).  We can also determine $B$ and 
$\gamma$ by considering the no penetration and zero temperature boundary conditions
on these 
expressions for $u_z^{(0)}$ and $\theta^{(0)}$. We find that both $B$ and $\gamma$ are purely imaginary:
\begin{gather}
B=\frac{-\mathrm{i}{\tilde k_x} Re (1+P)}{{R}_0^2\cosh\left(\sqrt{-{R}_0}/2\right)}\qquad \text{and} 
\qquad \gamma=\mathrm{i}{\tilde k_x} Re \left(\frac{1+P}{{R}_0^2} + 
\frac{1}{8{R}_0}\right). \label{eq:Bgam}
\end{gather}
With these expressions for $B$ and $\gamma$ we have acquired the complete expressions for 
$\omega_z^{(1)}$, $u_z^{(0)}$ and $\theta^{(0)}$.

Thus we now look at the next order of equation (\ref{eq:vortbc2}). At $\mathrm{O}(\epsilon^2)$ we find
\begin{equation}
\mathrm{i}{\tilde k_x} Re z\omega_z^{(1)} + {\tilde k_x}^2 + {\tilde k_y}^2
 -\frac{\mathrm{d}^2\omega_z^{(2)}}{\mathrm{d}z^2} = \frac{\mathrm{d}u_z^{(1)}}{\mathrm{d}z}.
\end{equation}
When taking the boundary conditions on the integral of this equation the final two terms will vanish since $\mathrm{d}\omega_z/\mathrm{d}z=0=u_z$ on the boundary. Therefore if we consider the integral of this equation
over the layer, substitute for $\omega_z^{(1)}$ and $B$ and $\gamma$ from equations (\ref{eq:omz1}) and (\ref{eq:Bgam}) respectively and apply the boundary conditions we obtain
\begin{gather}
\frac{1+P}{{R}_0^2} - \frac{2(1+P)\tanh\left(\sqrt{-{R}_0}/2\right)}{\sqrt{-{R}_0}{R}_0^2} + \frac{1+P}{12{R}_0} + 
\frac{1}{120} + \frac{1}{Re^2} \left(1 + \frac{k_y^2}{k_x^2} \right) = 0. \label{eq:R0solver}
\end{gather}
Equation (\ref{eq:R0solver}) can be solved numerically for $R_0$  using  given values of the parameters, $P$, $Re$ and $k_y/k_x$.

If we first consider the case $P=1$, $k_y/k_x=0$ and $Re=100$ we can compare the numeric results given by table \ref{tab:Ek} with those of 
table 4(b). We find that the asymptotics predict the numerics very well. For example at asymptotically small azimuthal wavenumber 
table \ref{tab:Ek} predicts that the Rayleigh number at onset will tend towards the value $-9.6577E^{-2}k_x^{-2}$. We see from table 4(b) 
that this gives excellent agreement. For modes with $k_y/k_x=0$ the asymptotics predict that $R_0$ is converging to approximately $-9.9$ with increasing zonal flow, which is also in excellent agreement with the numerics. Also of note is that equation (\ref{eq:R0solver}) has no negative $R_0$ solutions for $Re<10.9496$. As a result of this the asymptotic results, 
in  table 4(b), predict only modes with $R_0>0$ for $Re=10$. This is in excellent agreement with the numerics as the baroclinic modes were found to `switch-off' for approximately $Re<10.95$.

\begin{figure}[h]
\centering
\hspace{-60pt}
\includegraphics[width=0.40\textwidth]{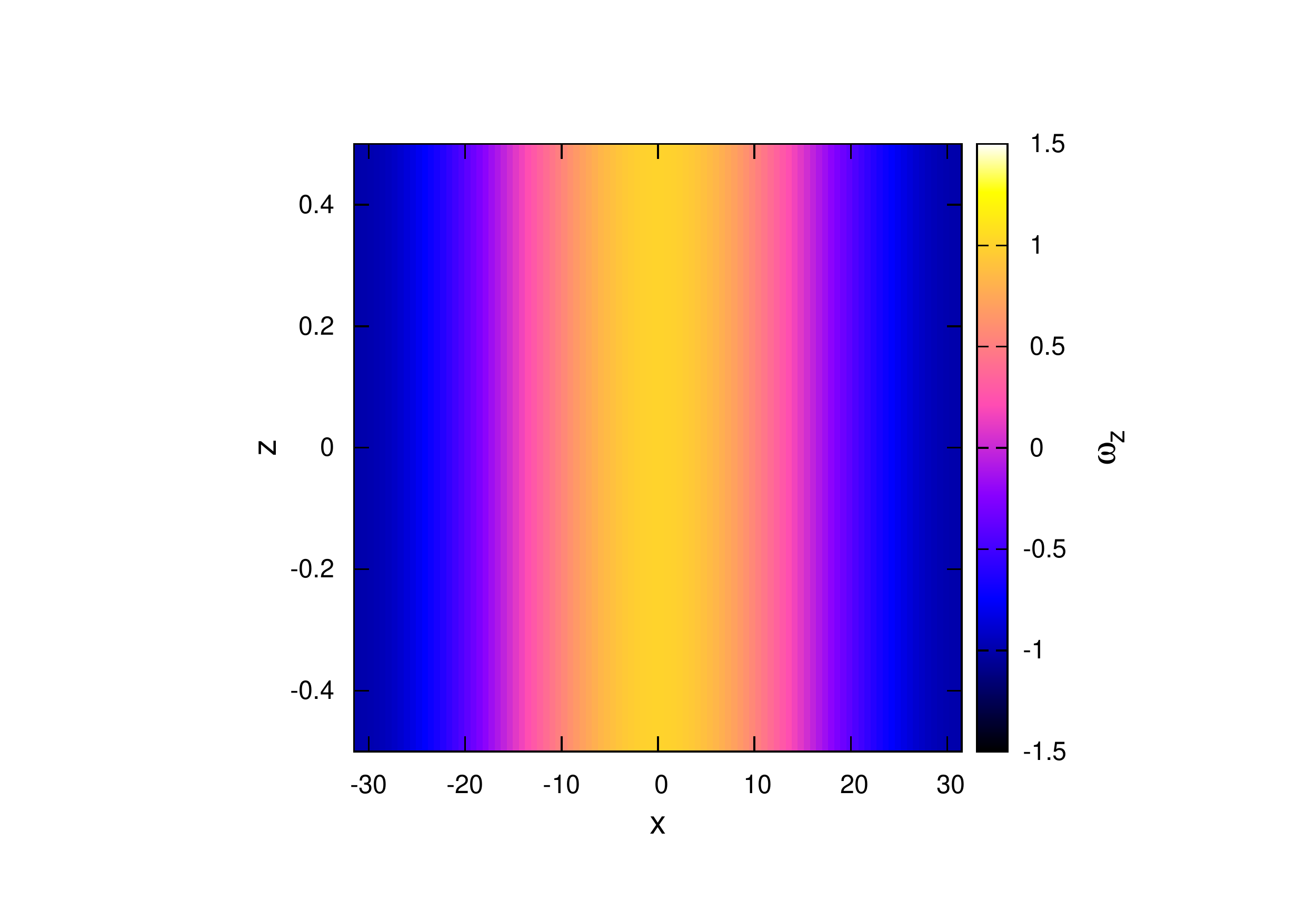}
\hspace{-60pt}
\includegraphics[width=0.40\textwidth]{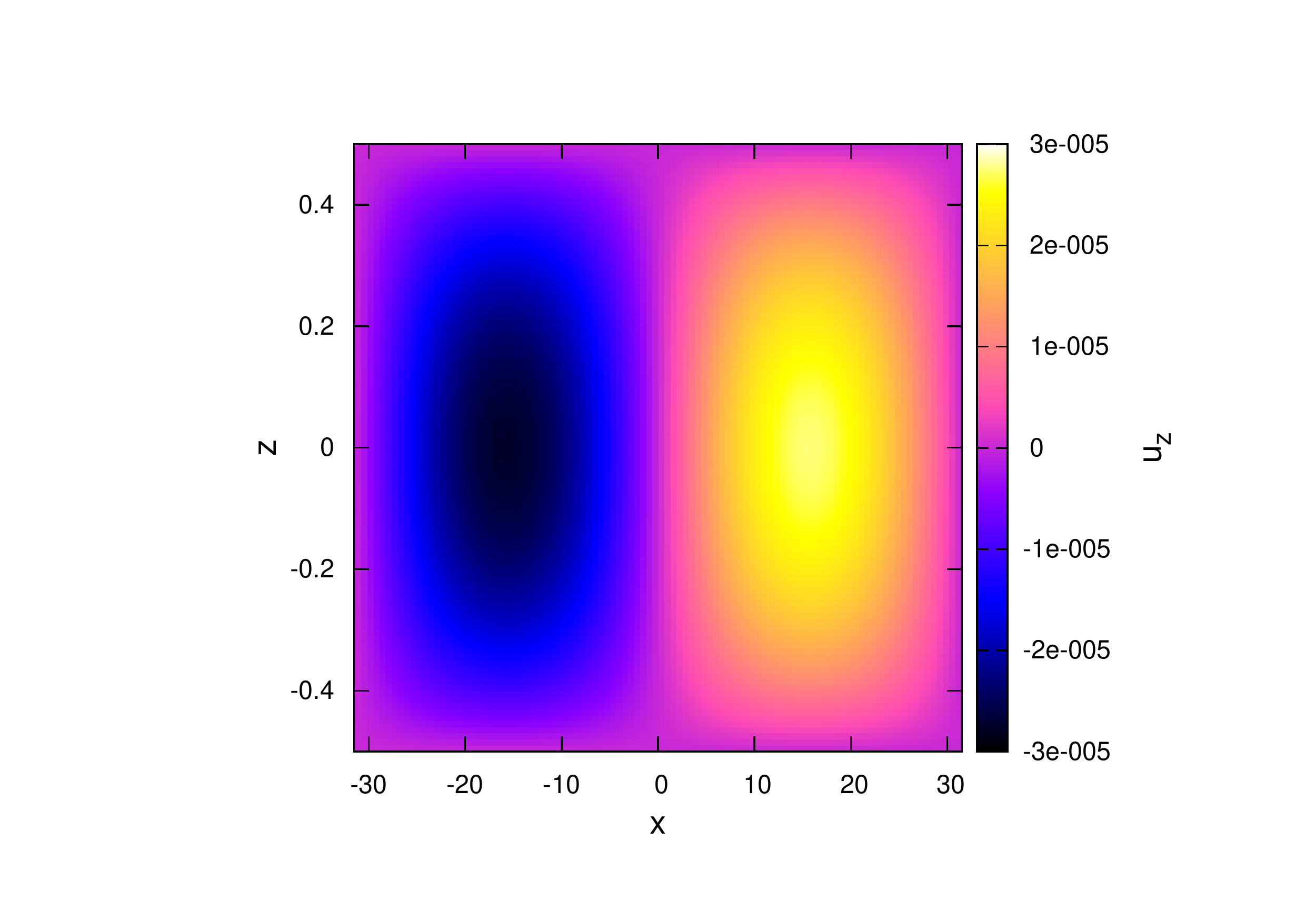}
\hspace{-60pt}
\includegraphics[width=0.40\textwidth]{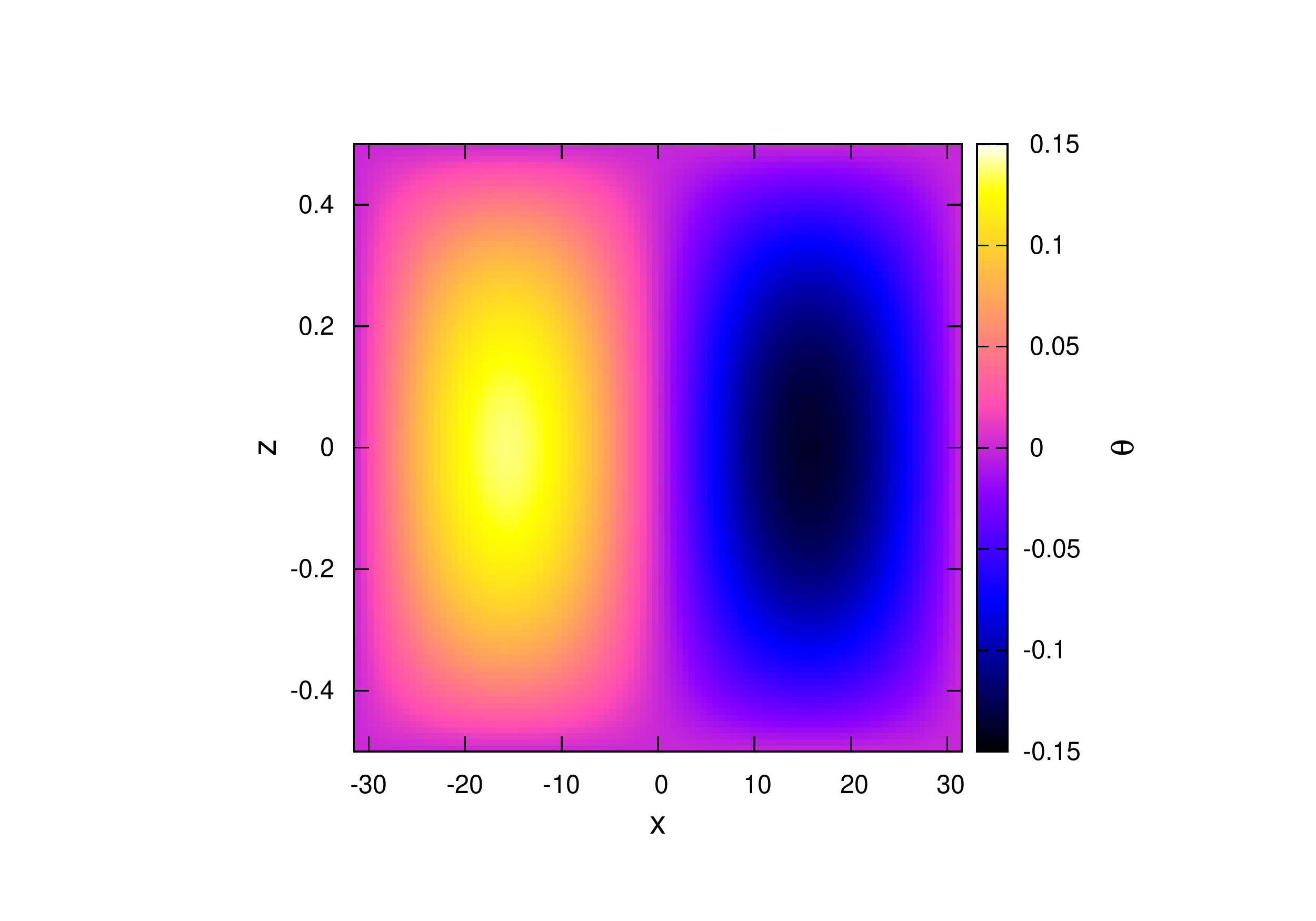}
\caption{Eigenfunction plots as predicted by the low wavenumber asymptotic theory. This is the equivalent of point $\times_2$ from figure \ref{fig:Raressf} where $E=10^{-4}$, $P=1$, $Ra=-10^6$, $Re=Re^*\equiv10.9599$, $k_x=0.1$ and $k_y=k_{y_c}=0$.}
\label{fig:asym}
\end{figure}

We can also see that the asymptotics of table 4(b) predict that increasing $k_y$ only serves to 
stabilise the system by increasing the Rayleigh number at onset in all cases. This matches the 
numerics as described in section \ref{sec:morenum}. In figure \ref{fig:asym} we have plotted the 
eigenfunctions predicted by the lowest order asymptotics as given by equations 
(\ref{eq:omznorm}), (\ref{eq:the0}) and (\ref{eq:uz0}) scaled using
$u_z= E {\tilde u_z}$, $\theta = E {\tilde \theta}$ in order to compare with
the equivalent parameter values at point $\times_2$ from figure \ref{fig:Raressf}. By comparing 
this plot with that of \ref{fig:x2} we can clearly see that the low wavenumber asymptotic theory is also predicting the correct form and magnitude of the eigenfunctions. The asymptotics continue to predict the correct form of the eigenfunctions for larger values of the Reynolds number where the onset parameter becomes the Rayleigh number, $Ra^*$.

\subsection{Relation to the Eady problem}

The low wavenumber equations (\ref{eq:vortbc2}) - (\ref{eq:tempbc2}) are related to the
quasi-geostrophic (QG) equations used in atmospheric science \citep[see e.g.][]{ped87}.
The geostrophic component of the velocity is given by $2 \Omega (u_x^G, u_y^G)
= (-\partial p / \partial y, \partial p / \partial x)$, so $\omega_z = - k^2 p / 2 \Omega$, 
and the pressure perturbation is simply proportional to the vertical vorticity, and
so equation (\ref{eq:curlvortbc2}) is simply the hydrostatic equation used in the QG approximation,
where vertical accelerations are neglected. The $y$-derivative terms in equations
(\ref{eq:vort1}) and (\ref{eq:heat2}) are also dropped in the Eady problem
\citep[see e.g.][p333]{drarei81} because of the low Rossby number assumption,
here $Re E << 1$. If we take the $z$-derivative of (\ref{eq:tempbc2}) and eliminate
$\tilde u_z$ and $\tilde \theta$ using (\ref{eq:vortbc2}) and (\ref{eq:curlvortbc2}),
we obtain
\begin{equation}
\left(\sigma P + \mathrm{i}k_x P Re z + k^2 - \frac{\mathrm{d}^2}{\mathrm{d}z^2}\right)\frac{\mathrm{d}^2 {\tilde \omega_z}}{\mathrm{d}z^2} +
\left(\sigma  + \mathrm{i}k_x  Re z + k^2 - \frac{\mathrm{d}^2}{\mathrm{d}z^2}\right) {\hat R}  {\tilde \omega_z} = 0.
 \label{eq:eady1}
\end{equation}
In the QG approximation, diffusion is usually ignored, and so the terms $k^2 - \mathrm{d}^2 / \mathrm{d}z^2$
are dropped in (\ref{eq:eady1}), leading to the classical Eady equation
\begin{equation}
\left(\sigma  + \mathrm{i}k_x  Re z \right)\left( \frac{\mathrm{d}^2 {\tilde \omega_z}}{\mathrm{d}z^2} + \frac{{\hat R}}{P} {\tilde \omega_z} \right) = 0,
 \label{eq:eady2}
\end{equation}
see e.g. equation (4.5.28) of \citet{drarei81}. The only boundary condition to
survive the neglect of diffusion is $\tilde u_z =0$, which leads to
\begin{equation}
\left(\sigma  + \mathrm{i}k_x  Re z \right)\frac{\mathrm{d} {\tilde \omega_z}}{\mathrm{d}z} =
 \mathrm{i}k_x P Re  {\tilde \omega_z} , \quad {\rm on} \quad z = \pm \frac{1}{2},
 \label{eq:eadybc}
\end{equation}
equivalent to equation (4.5.30) of \citet{drarei81}. Instability
occurs as an oscillatory mode, $\Im[\sigma] \ne 0$. The relevant part of our
parameter space is where $Re$ is large, since the viscosity is small, and there we
found oscillatory baroclinic modes as in figure \ref{fig:x3}, point $\times_
3$. 


\section{Conclusions}

The way in which convective instability and baroclinic instability interact in 
rapidly rotating systems has been elucidated. We found that the thermal
wind destabilises convective modes, lowering the critical Rayleigh number at which they onset. 
We also find that the critical azimuthal wavelength at onset lengthens. At a sufficiently
large Reynolds number, which in view of the very small viscosity occurring in many
geophysical systems can correspond to a rather small thermal wind, instability
becomes predominantly baroclinic, and the preferred azimuthal wavenumber tends to zero.
In our ideal plane layer geometry, there is no restriction on possible wavelengths,
but in more realistic spherical geometries, the boundaries will provide a limit. 
Slightly to our surprise, we found that convective modes and baroclinic modes are
smoothly connected, going through a transition region which can be studied
asymptotically (section 4.1) where the critical Rayleigh number smoothly
goes between positive and negative values. At the low azimuthal wavenumbers
preferred by baroclinic modes, an asymptotic analysis is possible (section 4.2) which
gives good agreement with the numerics in the stress-free case, and illuminates which
terms are important for instability. We also found that generally waves with non-zero
latitudinal wavenumber $k_y$ are not preferred in this problem, onset occurring in all cases
examined at the lowest $Ra$ when $k_y=0$.

At moderate Prandtl numbers, the onset of convection in this rotating B\'enard configuration
occurs with steady modes, but we find that at large Reynolds number oscillatory modes
are preferred. This result links our finite diffusion work with the quasi-geostrophic
shallow layer approximation used in atmospheric science, and in particular with the Eady problem
(section 4.3).

The existence of baroclinic instability in the physical conditions obtaining in planetary
interiors raises an interesting question of whether dynamo action could be driven 
by a heterogeneous core-mantle heat flux even if the core is stably stratified. 
This has also been investigated by \citet{sre09} where lateral variations were found to support 
a dynamo even when convection is weak.
It is widely believed that the heat flux passing from the Earth's core to its
mantle can vary by order one amounts with latitude and longitude, as a result
of cool slabs descending through the mantle and reaching the CMB from above. It is also generally believed
that the key criterion for the existence of a dynamo is that convection should be
occurring, and that the core is at least on average unstably stratified. However,
this analysis has raised the possibility that instabilities leading to fluid motion
driven by lateral temperature gradients can occur even when the fluid is strongly
stably stratified. Of course, it is not yet known whether the resulting nonlinear
motions would be suitable for driving a dynamo. In the plane layer geometry used
here, the preferred motion appears to be two-dimensional and therefore will
not drive a dynamo. However, in spherical geometry, and when secondary instabilities
may occur, dynamo action may become possible, in which case the view that convection
driven by an unstable temperature gradient is essential for dynamo action might 
have to be revised.

\bibliographystyle{gGAF}
\bibliography{paper7}

\end{document}